\documentclass[nofootinbib,aps,prd,reprint,superscriptaddress,floatfix]{revtex4-2}

\usepackage{listings}
\usepackage[utf8]{inputenc} 
\usepackage{graphicx,overpic,mathtools}
\usepackage{dsfont,amsthm,amsmath,amssymb,hyperref}
\usepackage{braket,bm,bbm,setspace}
\usepackage[normalem]{ulem} 
\usepackage{physics}
\usepackage{float}
\usepackage[makeroom]{cancel}
\usepackage[english]{babel}
\usepackage[table,xcdraw]{xcolor}
\usepackage{booktabs}
\usepackage{colortbl}
\usepackage{tensor}
\usepackage{verbatim}
\graphicspath{ {images/} }
\addto\captionsspanish{}
\hypersetup{
	colorlinks=true,
	pdfborder={0 0 0},
	citecolor=purple,
	linkcolor=blue,
	filecolor=blue,
	urlcolor=black,
}
\usepackage{subfigure}

\usepackage{bibunits}
\usepackage{orcidlink}

\makeatletter
\def\l@subsection#1#2{}
\def\l@subsubsection#1#2{}
\makeatother

\begin{document}
\title{A Quantum Signature Validation Algorithm \\ for Efficient Detection of Tampered Transactions in Blockchain}
\author{Jaime Torres \orcidlink{0009-0006-5146-9038}}
\email{jaitorre@ucm.es}
\author{Sergio A. Ortega \orcidlink{0000-0002-8237-7711}}
\email{sergioan@ucm.es}
\affiliation{Departamento de Física Teórica, Universidad Complutense de Madrid, 28040 Madrid, Spain}
\author{Miguel A. Martin-Delgado \orcidlink{0000-0003-2746-5062}}
\email{mardel@ucm.es}
\affiliation{Departamento de Física Teórica, Universidad Complutense de Madrid, 28040 Madrid, Spain}
\affiliation{CCS-Center for Computational Simulation, Universidad Politécnica de Madrid, 28660 Boadilla del Monte, Madrid, Spain.}

\begin{abstract}
{The Quantum Signature Validation Algorithm (QSVA) is introduced as a novel quantum-based approach designed to enhance the detection of tampered transactions in blockchain systems. Leveraging the powerful capabilities of quantum computing, especially within the framework of transaction-based blockchains, the QSVA aims to surpass classical methods in both speed and efficiency. By utilizing a quantum walk approach integrated with PageRank-based search algorithms, QSVA provides a robust mechanism for identifying fraudulent transactions. Our adaptation of the transaction graph representation efficiently verifies transactions by maintaining a current set of unspent transaction outputs (UTXOs) characteristic of models like Bitcoin. The QSVA not only amplifies detection efficacy through a quadratic speedup but also incorporates two competing quantum search algorithms---Quantum SearchRank and Randomized SearchRank---to explore their effectiveness as foundational components. Our results indicate that Randomized SearchRank, in particular, outperforms its counterpart in aligning with transaction rankings based on the Classical PageRank algorithm, ensuring more consistent detection probabilities.  These findings highlight the potential for quantum algorithms to revolutionize blockchain security by improving detection times to $O(\sqrt{N})$. Progress in Distributed Ledger Technologies (DLTs) could facilitate future integration of quantum solutions into more general distributed systems. As quantum technology continues to evolve, the QSVA stands as a promising strategy offering significant advancements in blockchain efficiency and security.}
\end{abstract}

\maketitle

\section{Introduction}\label{Introduction}

Blockchain technology was proposed in 2008 \cite{Bitcoin_nakamoto} as a decentralized system operating on a peer-to-peer (P2P) network, enabling direct transactions between participants without the need for centralized intermediaries or financial institutions. To eliminate reliance on a trusted central authority, transactions in the blockchain are validated through a distributed consensus mechanism within the P2P network. Additionally, cryptographic techniques ensure transaction security and privacy. Once transactions are validated, they become verifiable, immutable, and secure on the blockchain \cite{Chen2018blockchain}. This decentralized architecture strengthens the P2P model by ensuring transparency, distributing control, and mitigating single points of failure, positioning blockchain as a robust framework for trustless systems. The first operational blockchain system, based on the blockchain mechanism proposed in 2008, was launched in 2009 with the creation of Bitcoin. Since then, numerous blockchain systems have emerged, including Ethereum (2015) \cite{Ethereum}, Cardano (2017) \cite{Cardano}, and Polkadot (2020) \cite{Polkadot}. The rapid development of blockchain technology has driven innovation across diverse fields, such as medical data management \cite{Medical_data_blockchain}, logistics and supply chains \cite{Logistic_blockchain,Supply_chain_blockchain}, and art property management \cite{Art_blockchain}, among others. 

Despite its rapid deployment, blockchain faces critical challenges in scalability, efficiency, and security. These limitations have spurred interest in quantum computing as a transformative tool to enhance blockchain systems. Foundational results by Grover and Szegedy demonstrated the potential of quantum algorithms to outperform classical approaches in search and optimization tasks \cite{Grover,Szegedy}, motivating their application to blockchain. Recent advancements in quantum approaches to blockchain include the development of a highly efficient quantum consensus mechanism (QDPoS) \cite{Quantum_blockchain_2}, the application of the Quantum Approximate Optimization Algorithm (QAOA) to achieve cost-efficient logistics \cite{Quantum_blockchain_optimization}, and the implementation of Grover-based algorithms to optimize the block mining process \cite{Qauntum_blockchain_mining}. These advancements highlight the transformative potential of quantum computing in revolutionizing blockchain technology. In this context, we propose the Quantum Signature Validation Algorithm (QSVA), designed to enhance the detection of tampered transactions within blockchain systems, surpassing the performance of classical methods. The QSVA provides an efficient quantum-based search strategy for identifying fraudulent or impacted transactions in transaction-based blockchains. Leveraging the rapid progress in quantum computing, this model enables the early detection of manipulated transactions, significantly improving the efficiency, security, and reliability of blockchain technology. 

This paper is organized as follows. Section \ref{sec:Blockchain} provides a review of the fundamentals of blockchain technology, with a particular emphasis on its digital signature protocol and graph-based representations. Section \ref{sec:Algorithms} examines three key algorithms for blockchain transaction analysis: the Classical PageRank algorithm and two PageRank-based quantum search algorithms. In Section \ref{subsec:QSVA}, we present, step by step, the Quantum Signature Validation Algorithm (QSVA), designed for the efficient detection of tampered transactions. In Section \ref{sec:Simulation_results}, we evaluate the QSVA using a publicly available dataset of Bitcoin transactions, employing a quantum walk simulator executed on a classical computer. Furthermore, we assess whether the PageRank-based quantum search algorithms introduced earlier could serve as candidates for the QSVA by comparing their performance on the transaction dataset. Finally, Section \ref{sec:Conclusions} summarizes our findings and outlines future research directions.

\section{Blockchain technology}\label{sec:Blockchain}

In this section, we provide a brief review of the functioning of blockchain technology. To achieve this, we introduce the concepts of transaction blocks and digital signatures within the framework of cryptography. Throughout this paper, we exclusively focus on transaction-based blockchains. 

In blockchain technology, a distributed ledger (a record-keeping system that stores and validates transactions) is used to register entries across a peer-to-peer network. In public blockchains, this ledger is shared among all participants, and transaction information, such as the amount transferred or the transaction ID, is visible to anyone. However, the identities of the participants are typically pseudonymous, linked only to cryptographic addresses. This transparency allows operations on the network to be verified by multiple participants based on the publicly available information in the ledger, ensuring trust without relying on a central authority. Furthermore, the ledger is divided into blocks of cryptographically signed transactions. In turn, each block is cryptographically linked to the previous one, making it tamper-evident (designed to reveal any signs of unauthorized manipulation) \cite{Blockchain_overview}.

\subsection{Blockchain cryptography}\label{subsec:cryptography}

A digital signature may be used to detect whether or not the information was modified after it was signed. Every digital signature algorithm includes a signature generation process and a signature verification process. While a signatory uses the generation process to generate a digital signature on data, a verifier uses the verification process to verify the authenticity of each signature. The signature generation and verification processes employ a hash function \cite{Asymmetric_cryptography}.

The hash function takes a message or some data as entries and gives a fixed length string of bits. For example, the Bitcoin blockchain uses the SHA-256 hash function \cite{Bitcoin_nakamoto}. This means that, for (almost) every data given as an entry in the SHA-256, the function returns a string of 256 bits. Hash functions are widely used in blockchains due to their properties of preimage resistance and collision resistance \cite{Hash}. Preimage resistance ensures that, for nearly all specified outputs, it is computationally infeasible to find any input that produces that output when hashed. In contrast, collision resistance implies that it is computationally infeasible to find two distinct inputs that generate the same hash output. 

The combination of these two characteristics makes it computationally infeasible to obtain data or a message from a 256-specific output bit string. Consequently, the best way to match a hash output with its input is by ``trial and error'' or ``brute force''.
Nevertheless, even though trying by ``brute force'', the time required to find a specific output is excessive. Recall that there are $2^{256}$ different possibilities. Breaking the preimage resistance, an exhaustive search requires, on average, testing half of the total space of possible inputs, so we would have to try of the order of $5.79 \times 10^{76}$ guesses. A typical CPU for mining blocks in Bitcoin (in this paper, we will refer to the Bitcoin blockchain as ``Bitcoin''), which essentially is equivalent to trying data entries in an SHA-256 hash to obtain a particular output, can evaluate SHA-256 about 10 million times per second in the best case using an Intel Xeon Gold 6130 CPU at 2.10GHz \cite{Hash_brute_force}. This means that the time estimated to find the desired output would be around $1.84 \times 10^{62}$ CPU-years. Because of this, hash functions provide an effective way to encrypt information to avoid snoops who seek to know the information encrypted. 

\begin{figure}[hbtp]
    \centering
    \includegraphics[width=\columnwidth]{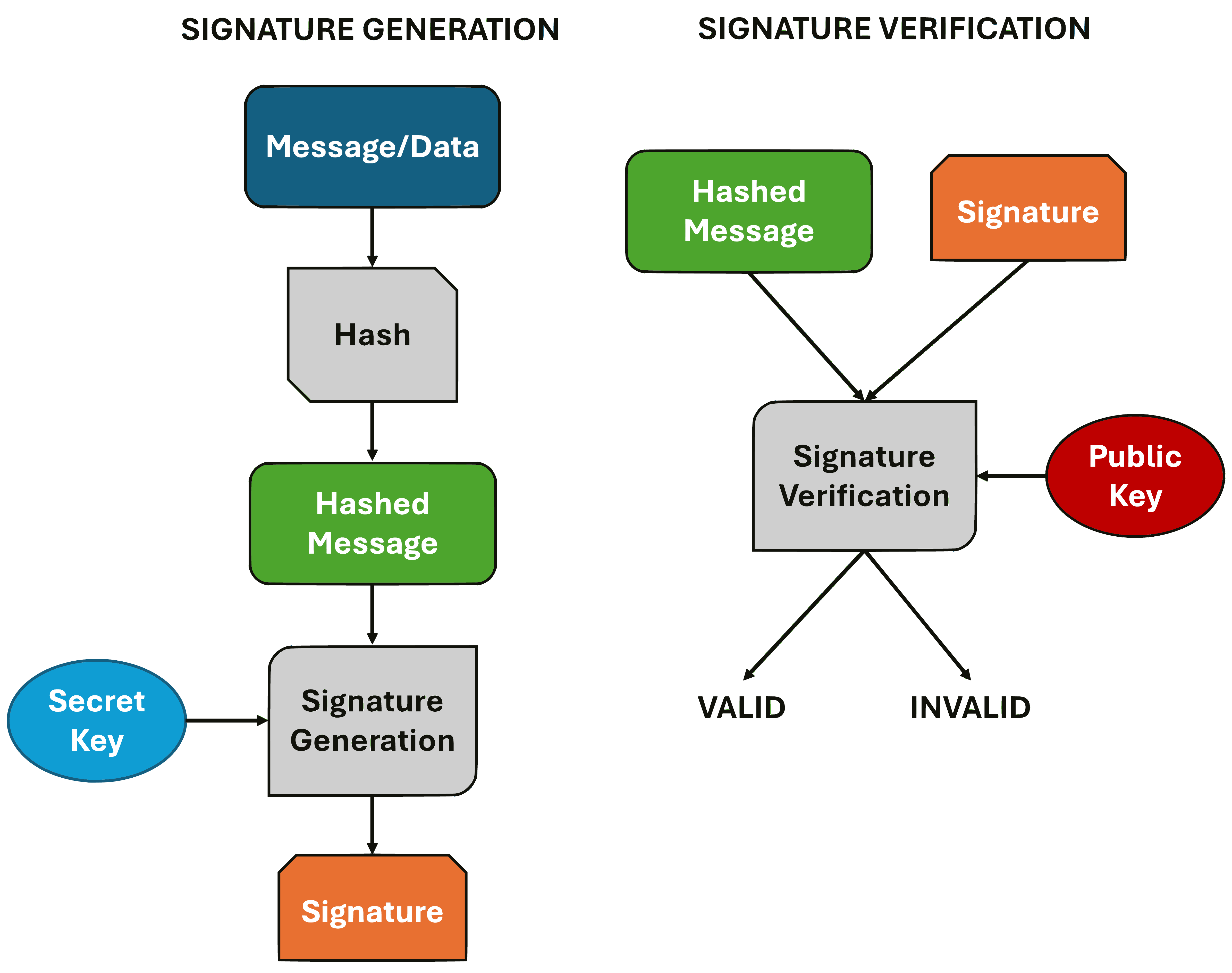}
    \caption{Digital Signature Processes: In the signature generation process, a hash function is applied to a message or data to obtain a message digest or hashed message. Then, a signature generation algorithm uses the hashed message and the signatory's secret key as inputs to create a digital signature. For signature verification, the hashed message and the digital signature generated during the signing process are used, along with the signatory's public key, in a signature verification algorithm to produce a boolean value indicating whether the message has been altered (invalid) or not (valid).}
    \label{fig:Digital_Signature}
\end{figure}

In Figure \ref{fig:Digital_Signature} we show the digital signature processes.

In addition to the hash function, a signatory must authenticate ownership of the digital signature to consolidate a signature. This is where the secret key and public key exchange takes place. Each signatory has a public and a secret key and is the owner of that key pair \cite{Asymmetric_cryptography}. Whereas the secret key can only be known by the signatory, that is, it must remain secret; the public key can be known by anyone who wants to verify a correctly signed message. After applying the hash function to a message, a signatory would use their secret key to generate a signature. Due to the collision resistance of the hash function, if the message is altered after it was signed, it would produce a different hash. This would cause the signature verification to fail when using the public key, revealing that the message has been modified after the signature was generated.
 
Formally, the processes involved in creating a digital signature constitute what is known as a signature scheme \cite{Digital_signatures}. Such a scheme typically comprises three algorithms: Gen, Sign, and Vrfy, along with an associated message space where the message lives. Each of these three algorithms can be regarded as mathematical functions, since they take inputs and produce outputs \cite{Ex_cryptography_1,Ex_cryptography_2}. In a digital signature, there is a function that we will call $P$, which takes as input a security parameter $k$ and outputs a pair of keys $(p_{k},s_{k})$, where $p_{k}$ corresponds to the public key and $s_{k}$ corresponds to the secret (or private) key. Additionally, there is a function, which we will call $S$, that takes as inputs the output of a hash function $\mathrm{H}$ applied to a message $m$ and the secret key generated from function $P$, and returns a signature $\sigma$. Finally, there is a function, which we will call $V$, that takes as inputs the output of a hash applied to a message, the signature associated with that message, and the public key generated by function $P$, which is linked to the secret key used by the signatory to sign the message. Function $V$ returns a boolean output. The output may be ``False'', in the case that the signature does not match the message, indicating that the message may have been altered after it was signed, or ``True'', otherwise. 

The functions associated with the three algorithms that substantiate the digital signature algorithm are listed in Table \ref{tab:Cryptograpic_functions}.

\begin{table}[hbtp]
    \caption{Functions involved in a digital signature process.}
    \centering
    \begin{tabular}{|c|c|c|}
    \hline
        \textbf{Function} & \textbf{Input} & \textbf{Output} \\ \hline
        $P$ & $k$ & $P$($k$) = ($p_{k}$,$s_{k}$) \\ \hline
        $S$ & $\mathrm{H}(m)$, $s_{k}$ & $S$($\mathrm{H}(m)$,$s_{k}$)= $\sigma$ \\ \hline
        $V$ & $\mathrm{H}(m)$, $\sigma$, $p_{k}$ & V($\mathrm{H}(m)$,$\sigma$,$p_{k}$) = 0/1 (T/F)\\ \hline
    \end{tabular}
    \label{tab:Cryptograpic_functions}
\end{table}

\subsection{Blocks of signed transactions}

So far, the distributed ledger of the blockchain is divided into blocks, each containing multiple transactions, with each transaction being digitally signed using the protocol and functions described in Section \ref{subsec:cryptography}. In terms of these functions, the hashed message $m$ corresponds to each transaction contained within a block of the blockchain.

\begin{figure*}[hbtp]
	\centering
        \includegraphics[scale=0.24]{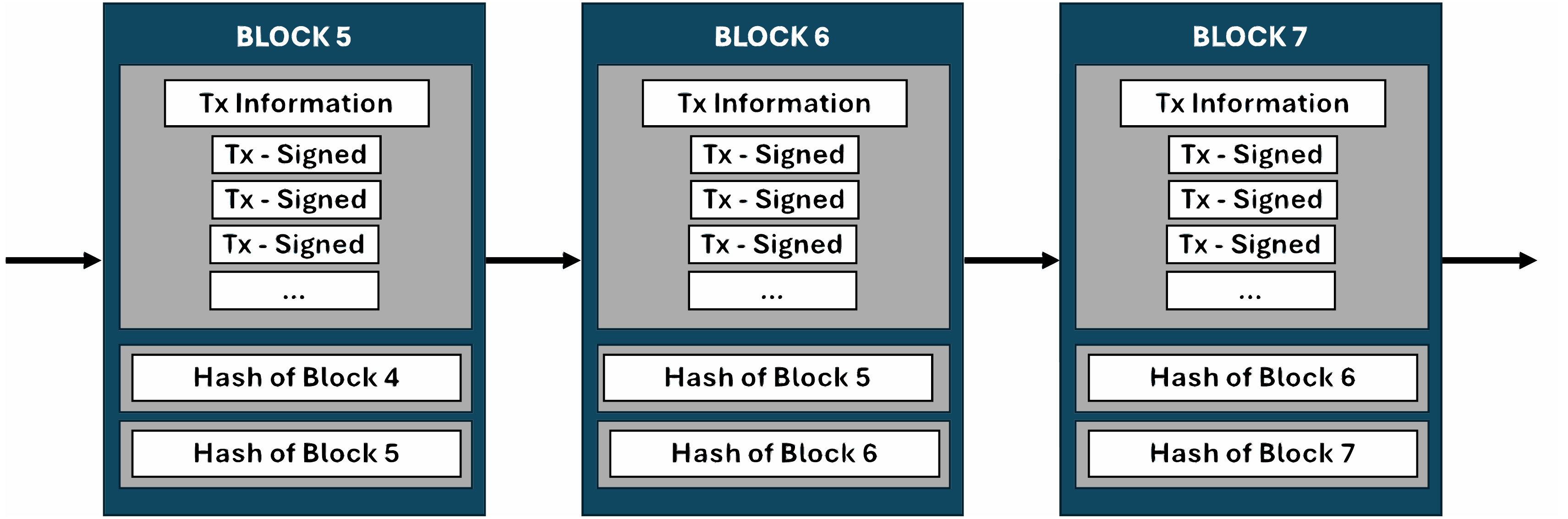}
        \caption{Schematic representation of the blockchain structure. Each block consists of a collection of digitally signed transactions, according to the protocol described in Section \ref{subsec:cryptography}, as well as the hash derived from the solved Proof-of-Work (PoW) required to add the block to the blockchain and the hash of the PoW from the preceding block.}
	\label{fig:Blockchain}
\end{figure*}

To incorporate new transactions conducted over the peer-to-peer network into the ledger, the system broadcasts each transaction to participants who maintain a personal copy of the ledger. From this point forward, we will refer to these participants as ``validators'', as they have the ability to verify the authenticity of transactions. Validators receive a list of digitally signed transactions resulting from the operations broadcast across the network. They store these transactions in a pending register known as the ``mempool'' prior to their verification. Validators then assess the validity of each transaction in the mempool by employing the verification function $V$, which takes the transaction as the hashed message, the signature attached to the transaction, and the available public key of the signatory. If the signature on the transaction fails to pass verification with the provided public key, this discrepancy indicates that the transaction (or message) may have been altered or tampered with, potentially by an attacker. In such cases, the validator removes the fraudulent transaction from its mempool, awaiting a valid one. 

Although there are various types of validators depending on the blockchain, the most well-known are those referred to as ``miners''. Miners validate signed transactions to compile a set of valid transactions and form a new block to be added to the blockchain. Upon successfully adding the block to the blockchain, miners receive an economic reward. 

Since validators determine which transactions are valid, the system employs a consensus protocol to prevent them from including fraudulent transactions in the blockchain. Therefore, the verification process is reliable as long as honest validators control the network \cite{Bitcoin_nakamoto}, meaning there are more honest validators than malicious ones. To prevent malicious validators from simulating multiple fake identities to take control of the consensus—a tactic known as a ``Sybil attack''—and to address other issues such as double-spending, a ``Proof-of-Work'' (PoW) system was proposed \cite{Bitcoin_nakamoto}. While other consensus mechanisms exist, we will focus solely on PoW-based blockchains. The PoW involves solving a computationally complex mathematical problem in order to add a new block to the blockchain, a process commonly referred to as ``mining''. In Bitcoin, for example, the PoW process utilizes a hash function. This process requires significant computational power, which increases with the number of simulated identities an attacker uses to gain control of the network. Additionally, to ensure greater security, each block in the blockchain is linked to the previous one by the hash associated with the PoW, as shown in Figure \ref{fig:Blockchain}. This means that an attacker would have to redo all the computational work of the blocks that follow the altered block to benefit from their modifications. 

\begin{figure}[hbtp]
    \centering
    \includegraphics[width=\columnwidth]{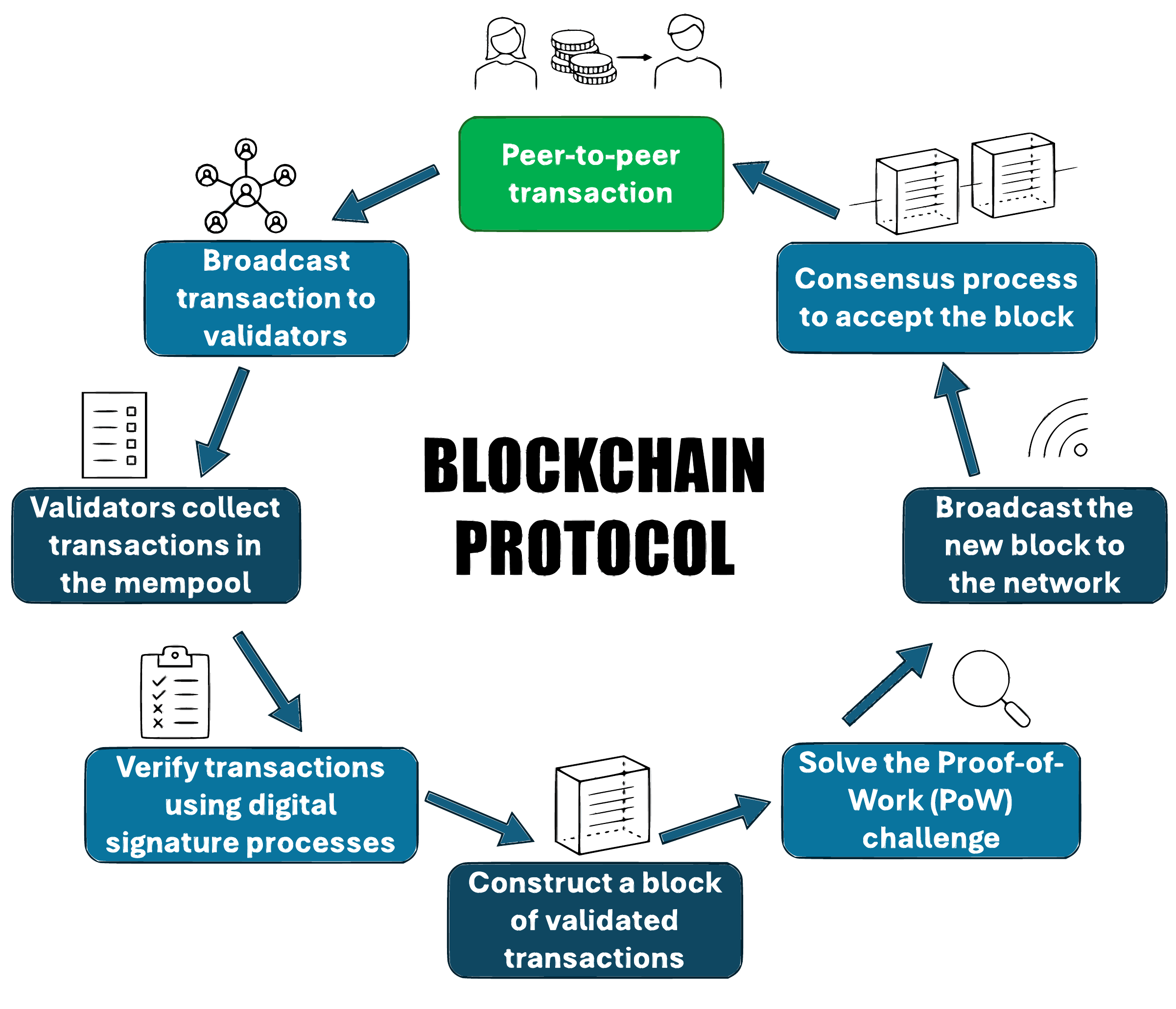}
    \caption{Schematic representation of the blockchain protocol. The starting point is highlighted in green. The diagram illustrates the sequential steps involved in the protocol, including transaction validation, block formation, Proof-of-Work mechanism, and block propagation across the network. Each step ensures the integrity, security, and decentralization of the distributed ledger system.}
    \label{fig:Blockchain_protocol}
\end{figure}

After this discussion, we can develop an intuitive understanding of how blockchain technology operates: each new transaction on the peer-to-peer network is broadcast to all validators. Each validator maintains a copy of the ledger, which is organized into blocks containing signed transactions. Validators collect incoming transactions in the mempool, verify their validity using a digital signature verification process, and construct a block of validated transactions by solving the Proof-of-Work (PoW) challenge. The newly constructed block is then broadcast to the network, where it is accepted through a consensus process. Validators then proceed to work on building the next block in the chain, incorporating the hash of the accepted block as the previous hash in the new block under construction, see Figure \ref{fig:Blockchain_protocol}.

\subsection{Graph representation}\label{subsec:graph_representation}

\begin{figure*}[hbtp]
	\centering
        \subfigure[]{\includegraphics[scale=0.13]{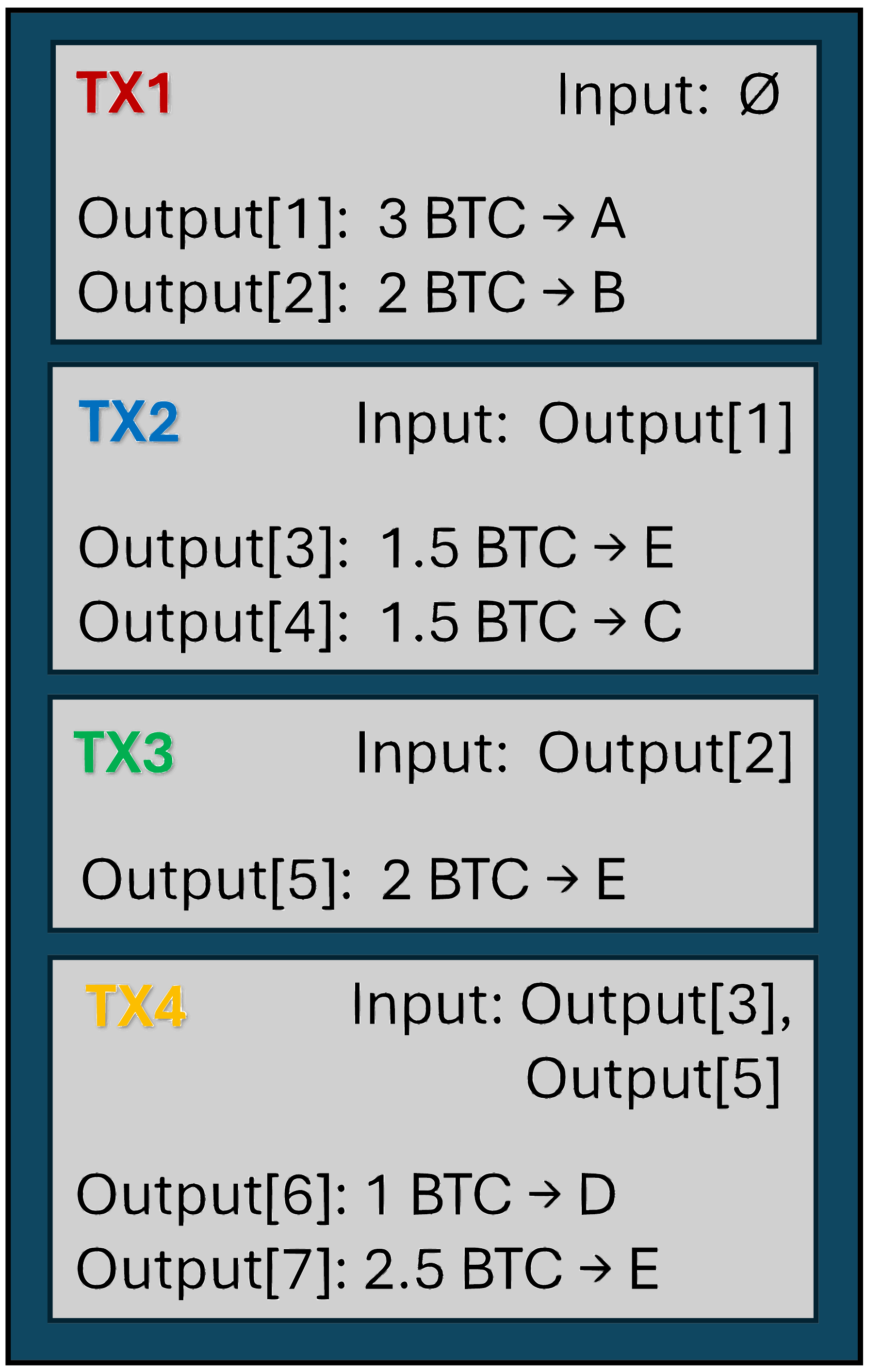}\label{fig:UTXO_ledger}}
        \hspace{10mm}
	\subfigure[]{\includegraphics[scale=0.12]{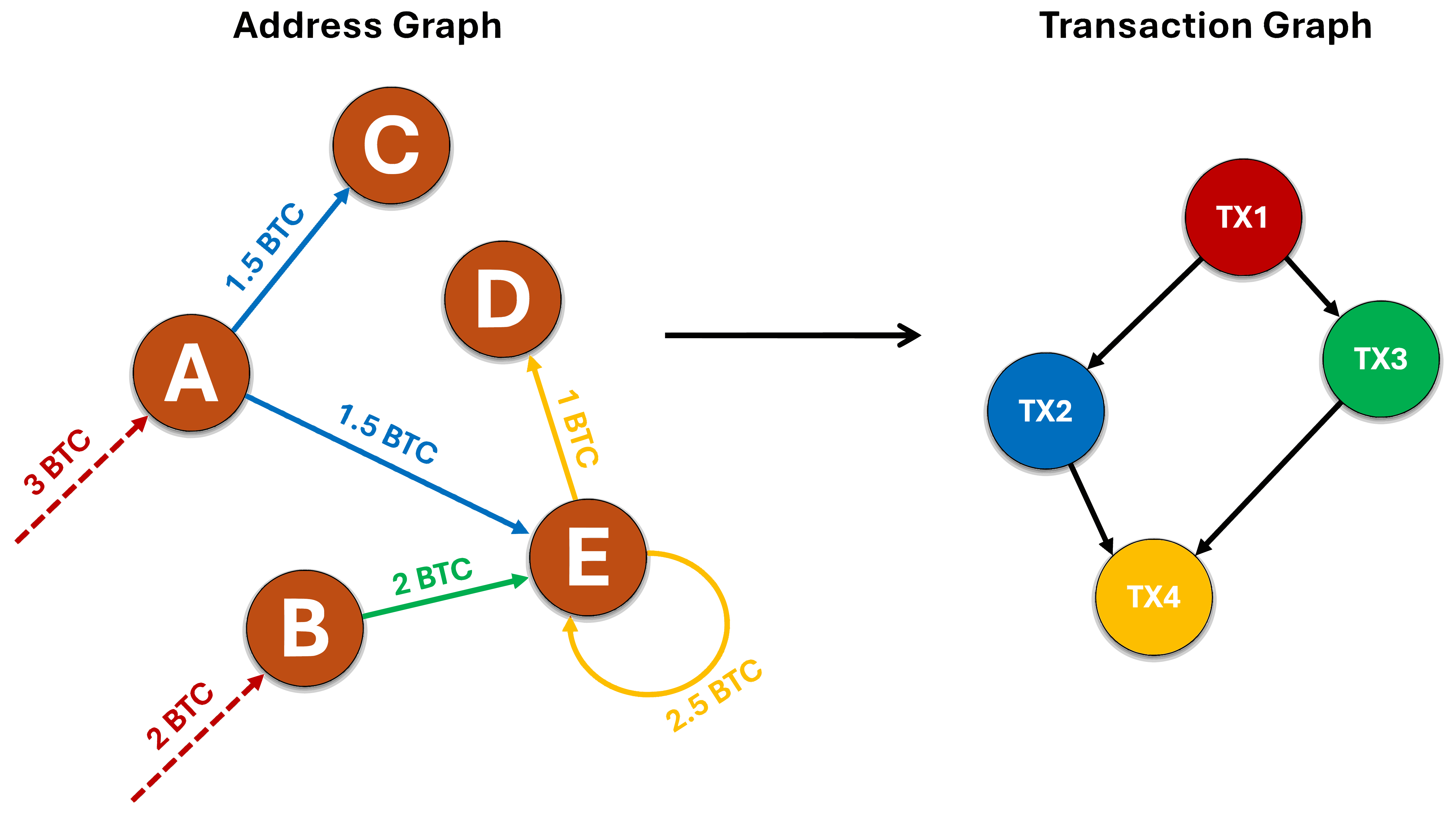}\label{fig:UTXO_graph}}
    \caption{a) Schematic representation of Bitcoin's UTXO-based ledger. Outputs specify the amounts of BTC sent to specific addresses, while inputs reference the outputs of previous transactions. If an input is empty, it indicates the creation of new BTC, such as through the mining process of a block. b) Example of an address graph and its associated transaction graph. The transaction graph can be constructed from the UTXO-based ledger following the process described in Section \ref{subsec:graph_representation}.}
	\label{fig:UTXO}
\end{figure*}

Since beneath the blockchain lies a peer-to-peer network, an effective approach to extracting useful information from it is to represent the network as a graph. Various graph models can be found in the literature, each tailored to the specific type of information of interest \cite{Cryptocurrency_overview}. Given that our study focuses on analyzing individual transactions, we will use the transaction graph representation. In this representation, transactions are treated as nodes of a graph, and the flow of available funds transferred between transactions is represented as directed edges. Unlike the address representation, where funds are transferred from a source address to a destination address, this model represents the movement of assets from a source transaction to a destination transaction.

This transaction-centered model stems naturally from the Bitcoin data structure, where it is called UTXO (Unspent Transaction Output) \cite{UTXO}. In Bitcoin, each transaction consists of two main parts: an output, which contains an amount of BTC—defined as the digital currency exchanged in Bitcoin transactions—and a script that specifies the conditions under which the funds can be spent, effectively representing coins available for future transactions; and an input, which is an unambiguous reference to the outputs of previous transactions and serves as proof that the spender has the right to use those funds, see Figure \ref{fig:UTXO_ledger}. The UTXO model has been extensively discussed and analyzed in prior works \cite{Didactic_UTXO,Formal_UTXO}. This model enables efficient transaction verification, as it only requires checking the transaction output to confirm that the amount to be transferred is available and unspent. Unlike account-based systems, there is no need to track all past transactions of an address from the beginning of the blockchain to determine its balance. Instead, it is enough to verify that the referenced output in the transaction remains unspent \cite{Didactic_UTXO}. In the UTXO scheme, a transaction output either is consumed in its entirety by a transaction, or none of it. This ensures that the total value of BTC in each transaction is preserved, a principle that we refer to as the ``conservation rule'', meaning that the sum of the inputs equals the sum of the outputs (plus any change returned to the sender).

The construction of the transaction graph is directly derived from a UTXO-based model, such as Bitcoin. Each transaction in the UTXO model is associated with a node. Nodes are connected by directed edges that point from the previous transaction, whose output has not been spent, to the transaction that receives the funds and references the unspent output in its input. Therefore, a transaction graph, similar to the one shown in Figure \ref{fig:UTXO_graph}, can be constructed from the transactions in the UTXO model. Moreover, in transaction graphs, the flow of BTC between nodes is preserved.

Transaction graphs are widely used for multiple purposes, such as analyzing money flows \cite{Transaction_graph_analysis}, detecting money laundering transactions \cite{Money_laundering}, and conducting typical network graph analyses, such as determining the clustering coefficient or applying algorithms like PageRank \cite{PageRank_transaction_graph}. In the next section, we will introduce graph-based algorithms, including quantum algorithms, that can be applied to the Bitcoin transaction graph to analyze tampered transactions.

\section{Algorithmic analysis of a Bitcoin transaction graph}\label{sec:Algorithms}

In this section, we first introduce a classical algorithm extensively applied in the analysis of social networks, transportation networks, and search engines: the so called PageRank algorithm \cite{Brin1,Brin2,Brin3}. Subsequently, we turn our attention to quantum search algorithms, which have demonstrated superior performance compared to their classical counterparts. We begin by discussing a foundational algorithm of paramount importance in applications such as databases, text processing, and artificial intelligence: the Grover algorithm \cite{Grover}.

We then present two graph-based alternatives to the Grover algorithm found in the literature: the Quantum SearchRank \cite{Searchrank} and the Randomized SearchRank \cite{Randomized} algorithms. By leveraging their graph-based approach, these algorithms effectively exploit the structure of Bitcoin’s transaction graph, facilitating their implementation on the blockchain. Furthermore, their PageRank-driven nature enables a more efficient search strategy for the early detection of manipulated transactions within the blockchain, potentially offering advantages over the original Grover algorithm, as discussed in the following sections.

This motivates the development of the Quantum Signature Validation Algorithm, a quantum algorithm designed for the early detection of such fraudulent activities within the blockchain.

\subsection{Classical PageRank}\label{subsec:Classical_PageRank}

PageRank was first introduced in 1998 \cite{Brin1} as an algorithm to help establish objective criteria for ranking web pages at a time when the dominating ranking of web pages was carried out using subjective criteria in search engines. Google was the first prototype that used the PageRank algorithm, and its search criteria has proven, to this day, to be highly effective compared to other search engines based on opinion-based rankings \cite{Brin3}. 

The key idea behind Google's PageRank algorithm is that the importance of a page is determined not only by the number of pages linking to it but also by the quality of these links. Specifically, pages with incoming links from pages that have few outgoing links contribute more to the target page's importance. As a result, pages that receive many incoming links from relevant sources are more likely to rank higher in PageRank. As we will explain in Section \ref{subsec:QSVA}, this approach could be applied to blockchain transaction analysis to efficiently identify tampered transactions.

PageRank algorithm consists in a random walk on a graph associated with a Markov chain, which is represented as a column-stochastic matrix constructed the following way. First, we will define the matrix elements of a non-stochastic matrix as follows: 
\begin{equation}\label{eq:Hyperlink_matrix}
	H_{i,j} := 
	\left\lbrace\begin{array}{c}
		1/\text{outdeg}(P_j) \ \ \ \text{if} \ P_j \in B_i,\\
		0 \ \ \ \ \ \ \ \ \ \ \ \ \ \ \ \ \ \ \ \text{otherwise},
	\end{array}
	\right.
\end{equation} 
where outdeg$(P_{j})$ is the outdegree of the page $P_{j}$ (or the node $j$ of the graph associated with the Markov chain), and $B_{i}$ is a set of pages linking to it.

From matrix $H$ (called ``Hyperlink matrix''), we can build the column-stochastic matrix $E$ by substituting the column corresponding to a dangling node with a column of all 1/N, with N the number of nodes \cite{Paparo1}. Finally, the matrix that represents the graph associated with the Markov chain where the random walk takes place, the so-called Google matrix, is obtained through the following expression:
\begin{equation}
\label{eq:Google_matrix}
    G:=\alpha E + \frac{(1-\alpha)}{N}\textbf{1},
\end{equation}
where $\textbf{1}$ is a matrix in which all entries are equal to $1$. In this expression, $\alpha$ is a free parameter called ``damping parameter'', whose values lie in [0,1]. For the PageRank algorithm, $\alpha=0.85$ is chosen, as this value optimizes the performance of the algorithm. The matrix elements, $G_{ij}$, represent the probability of going from the node $j$ to the node $i$ of the graph associated with the matrix. The goal of the algorithm is to obtain the eigenvector with eigenvalue 1. The power method is an effective way to achieve it due to the size of the matrix. This method typically involves multiplying the Google matrix by an initial uniform vector $I_{0}$, which ensures that all the pages have the same importance, and then multiply re-iteratively the Google matrix by the resulting vectors until the stationary state $I_{st}$ is reached, that is $I_{st}=GI_{st}$. This process is equivalent to performing a random walk on the graph the Google matrix represents until it converges:
\begin{equation}\label{eq:Stationary_state_PageRank}
    I_{st}=\lim_{t \to \infty}G^{t}I_{0}, \hspace{3mm}  t\in \mathbb{N}.
\end{equation}
The stationary state corresponds to the eigenvector associated with the eigenvalue 1. Its components represent the pages of the web, and its values represent the probability that a user ends up on a given web page after the random walk. Thus, the more important pages in PageRank have a greater associated probability, providing an objective ranking of pages.

\subsection{Quantum SearchRank}\label{subsec:Quantum_SearchRank}
 
Generally, a search algorithm is a set of instructions whose goal is to find a specific element in a database through a number of queries. Supposing a database has $N$ elements, a classical search algorithm would need at least $O(N)$ queries to find the desired element. That is, we have to check for every element in the database to see if it is the searched element. This process continues until the desired element is identified. Surprisingly, quantum properties allow reducing the number of queries required to find the desired element to $O(\sqrt{N})$ using quantum computing. Grover was the first to notice this speed-up while using quantum superposition and interference in the elements of a database \cite{Grover}. A more general algorithm to find multiple desired elements in a database can be achieved using an extension of the Grover algorithm \cite{Portugal, Grover_M}. In this case, if $M$ are the elements that we want to find and $N$ is the total number of elements in the database, the minimum number of queries to find any of the $M$ desired elements would be $O(\sqrt{N/M})$, in contrast to its classical equivalent which would need $O({N/M})$. Although an algorithm similar to Grover's could be implemented for use in Bitcoin, we will explore an equivalent graph-based algorithm that naturally fits Bitcoin's transaction graph: the Quantum SearchRank.

Quantum SearchRank is a graph-based quantum algorithm proposed in 2014 \cite{Searchrank} as a combination of the Grover algorithm and the Quantum PageRank algorithm \cite{Paparo1,Paparo2,APR}. The probability of measuring a marked node is not only amplified but also depends on the PageRank distribution. The goal of this algorithm is therefore to identify and rank a set of database elements in a manner similar to the PageRank algorithm. Given this approach, we refer to such quantum search algorithms as PageRank-based quantum search algorithms throughout this paper. At its core, the algorithm relies on a quantum walk over a quantized Markov chain proposed by Szegedy \cite{Szegedy}. 

In the Quantum SearchRank algorithm, the Google matrix \eqref{eq:Google_matrix} serves as an $N \times N$ stochastic matrix that defines a Markov chain on a graph with $N$ nodes, on which the quantum walk is performed. Each node of the graph represents an element of a $N$-element database. The damping parameter is set to $\alpha=0.25$ in the Quantum SearchRank algorithm \cite{Searchrank}, compared to the $\alpha=0.85$ employed in the Classical PageRank algorithm. The Hilbert space where the quantum walk takes place is spanned by all the vectors that represent the $N \times N$ directed edges of the graph \cite{Paparo2}, more precisely, $\mathcal{H}= \text{span} \{\ket{i}_{1}\ket{j}_{2}, \hspace{0.3cm} i,j \in N \times N \} = \mathbb{C}^{N}$ $\otimes$ $\mathbb{C}^{N} $. The states with indexes 1 and 2 refer to the nodes on two copies of the original graph. We will count the nodes of the graph and the matrix indices associated with the nodes from 0 to $N-1$.

To take advantage of the quantum superposition, we define the following states:
\begin{equation}\label{eq:psi_i}
	\left|\psi_i\right> := \left|i\right>_1 \otimes \sum_{k=0}^{N-1} \sqrt{G_{ki}}\left|k\right>_2,
\end{equation}
which are a superposition of the vectors representing the edges outgoing from the $i^{th}$ node.

To perform the quantum walk, we have to define the unitary evolution operator \cite{Notes}. First, let us define the projector operator onto the subspace generated by the $\left|\psi_i\right>$ states:
\begin{equation}
    \Pi := \sum_{i=0}^{N-1}\left|\psi_i\right>\left<\psi_i\right|.
\end{equation}
A reflection operator $R$ over this subspace can be also defined:
\begin{equation}
    R:=2\Pi-\mathds{1}.
\end{equation}
We introduce a swap operator that interchanges the states between the two quantum registers:
\begin{equation}
    S:=\sum_{i,j=0}^{N-1}\left|i,j\right>\left< j,i\right|.
\end{equation}
To identify any of the $M$ searched elements, we will introduce a quantum operator $Q_{f}$, called oracle.
The oracle acts through a boolean function $f(x)$ that takes a bit string of $n$ bits as input, and returns a boolean output:
\begin{equation}\label{eq:f}
    f(x):\{0,1\}^n\xrightarrow{}\{0,1\}.
\end{equation}

From a quantum circuit perspective, the oracle is a black box operator that acts on a composite state of a data register and a target register as follows:
\begin{equation}\label{eq:oracle_circuit}
    Q_{f}\left|x\right>\left|y\right>=\left|x\right>\left|y \oplus f(x)\right>,
\end{equation}
being $\oplus$ the bitwise XOR operator. When the target is in state $\left|0\right>$, it just stores what the function $f$ returns when applied to the register data, which returns $1$ if $x$ is a searched element and $0$ otherwise. The index $x$ of the data resister can be used as a reference to a database. Thus, $f(x)$ can be modeled to retrieve data from a database associated with the index $x$ and perform calculations based on that data. 

From the graph's states perspective, we will define the action of the oracle to be,
\begin{equation}\label{eq:oracle_graph}
	Q_{f}\left|i\right> := 
	\left\lbrace\begin{array}{c}
		-\left|i\right> \ \ \ \ \ \text{if} \ i \in \mathcal{M},\\
		\ \ \ \ \left|i\right> \ \ \ \ \    \text{otherwise},
	\end{array}
	\right.
\end{equation}
where $\mathcal{M}$ is a set which contains the $M$ desired elements. Note that the action of the oracle on a searched element changes its sign, effectively marking that element for identification. This marking process is why these elements are referred to as ``marked elements'', or ``marked nodes''. The marking is achieved by initializing the target qubit in state $\left|-\right> = (\left|0\right>-\left|1\right>)/\sqrt{2}$. When $f(x)=1$, the addition leaves the target qubit in $-\left|-\right>$. Nevertheless, the negative sign can be attributed to the overall system as a global phase, allowing the target qubit to be traced out of the system. In the SearchRank algorithm, the oracle acts on the first register, so we define the SearchRank oracle as follows:
\begin{equation}
    Q_{1}:=Q_f \otimes \mathds{1}.
\end{equation}

The unitary evolution operator can be constructed using the previous operators:
\begin{equation}\label{eq:unitary_evolution}
    W_{Q}=\left(SQ_{1}R\right)^{2}.
\end{equation}
Due to the swap operator that it contains, the unitary evolution operator has to be defined with an even power \cite{Randomized}. The Quantum SearchRank algorithm is carried out by applying the unitary evolution operator to an initial state, formed as an equal superposition of all the $\left|\psi_i\right>$ states, to perform the quantum walk:
\begin{equation}\label{initial}
	\left|\Psi^{(0)}\right> := \frac{1}{\sqrt{N}} \sum_{i=0}^{N-1} \left|\psi_i\right>.
\end{equation}
Thanks to quantum parallelism, we can apply the function $f(x)$ to all states $\left|\psi_i\right>$ simultaneously through the action of the oracle, which is contained in the unitary evolution operator, on the superposition state $\left|\Psi^{(0)}\right>$:
\begin{equation}\label{initial}
	\left|\Psi^{(t)}\right> = W_{Q}^{t}\left|\Psi^{(0)}\right>.
\end{equation}
Note the the similarities between this expression and \eqref{eq:Stationary_state_PageRank}.

Finally, we measure the second register to obtain the instantaneous SearchRank distribution at each time step $t$:
\begin{equation}
	S_q(P_i,t) := \left|\left|\tensor[_2]{\big<i\left|\Psi^{(t)}\right>}{}\right|\right|^2.
\end{equation}
But due to the unitary evolution, this probability exhibits oscillations depending on the value of $t$. Therefore, it is crucial to measure the system at an specific step $t_Q$, which guarantees that the probability of measuring a marked node reaches its maximum value. Quantum interference allows obtaining the maximum probability when the system is measured at $t_Q\approx\sqrt{N/M}$ \cite{Searchrank}. As a result, any of the marked elements in $\mathcal{M}$ can be obtained with high probability after measurement, using $O(\sqrt{N/M})$ queries.

\subsection{Randomized SearchRank}\label{subsec:Randomized_SearchRank}

An innovative algorithm, based on the SearchRank algorithm and semiclassical walks \cite{Semiclassical}, has been proposed \cite{Randomized}. It is called ``Randomized SearchRank'' because, instead of employing an initial state of equal superposition such as $\left|\Psi^{(0)}\right>$, the system is initialized in a mixed state:
\begin{equation}\label{rho}
	\rho := \frac{1}{N} \sum_{i=0}^{N-1} \left|\psi_i\right>\left<\psi_i\right|.
\end{equation}
The key idea is to take advantage not only of quantum superposition but also of the statistical combination of quantum states simultaneously through the use of a mixed state. After initialization, the algorithm is similar to the Quantum SearchRank: the unitary evolution operator, $W_Q$, must be applied to the initial mixed state, $\rho$, a number of times, $t$, to perform the quantum walk. After the quantum walk, the second register is measured to obtain the probability of measuring a marked node,
\begin{eqnarray}\label{trace_1}
R_q(P_j,t) :&=& \text{Tr}_1 \left[\tensor[_2]{\left<j\right|W_Q^{t} \rho (W_Q^{t})^\dagger\left|j\right>}{_2}\right]\nonumber\\
&=&\sum_{i=0}^{N-1}\frac{1}{N}\left|\left|\tensor[_2]{\left<j\right|W_Q^{t}\left|\psi_i\right>}{}\right|\right|^2.
\end{eqnarray}
As in the case of the Quantum SearchRank algorithm, the probability oscillates as a function of the number of steps $t$ for which the quantum walk is performed. Therefore, the system must be measured at a specific step $t_R$ that ensures that the probability of measuring a marked node is maximized. The minimum number of times, $t_R$, that the unitary evolution operator must act on the mixed state to guarantee that any of the $M$ marked elements is obtained with high probability after measurement is also $t_R \approx \sqrt{N/M}$ \cite{Randomized}. 

It is important to emphasize that the probability of measuring a marked node at the maximum drops with $N/M$ for the Quantum SearchRank, whereas it remains close to 1 for any value of the ratio $N/M$ for the Randomized SearchRank \cite{Randomized}. This is particularly noticeable for big networks, resulting in a malfunction of the Quantum SearchRank in acquiring marked nodes post-measurement. Therefore, Randomized SearchRank could be a great alternative for a wide variety of networks.

\subsection{Quantum Signature Validation Algorithm}\label{subsec:QSVA}

Let us recap some of the points covered so far. On the one hand, in Section \ref{sec:Blockchain}, we examined how a blockchain system was introduced to eliminate the need for trusting a third party to execute transactions. In this system, transactions are recorded in a ledger divided into blocks and digitally signed using cryptographic algorithms, which can be represented as mathematical functions. New transactions are broadcast to participants with access to the ledger, known as validators. Validators collect these transactions in a mempool and verify that they have not been tampered with after being signed. Upon collecting a set of valid transactions from the mempool, validators assemble a block of transactions and perform computational work to solve the PoW challenge, thereby appending the block to the blockchain. Additionally, we highlighted that in the Bitcoin blockchain, transactions are recorded following a UTXO model, which has a direct representation as a graph. On the other hand, in Section \ref{subsec:Quantum_SearchRank} and \ref{subsec:Randomized_SearchRank}, we explored two graph-based quantum algorithms that significantly accelerate the search for an element within a database compared to their classical counterparts, the Quantum SearchRank and the Randomized SearchRank. 

It is clear that early detection of tampered transactions is crucial, as such transactions would be removed when discarded promptly from the validators' mempool. The faster this occurs, the sooner valid transactions can be included in a block, added to the blockchain, and executed. This observation, combined with the rapid search capabilities of quantum algorithms, motivates their application in processing transactions within a blockchain, thereby enhancing the overall efficiency of the blockchain system. Furthermore, the UTXO structure of transactions in blockchains such as Bitcoin simplifies the implementation of graph-based algorithms by leveraging the transaction graph. Building on this concept, we propose a graph-based quantum algorithm intended for implementation on a quantum computer. This algorithm aims to enable the rapid detection of tampered transactions, thereby contributing to the optimization of blockchain systems such as Bitcoin. Since the core of the algorithm involves the process of transaction validation, the algorithm is named ``Quantum Signature Validation Algorithm'' (QSVA). In this section, we examine the details related to the deployment of the algorithm.

The first fact to note is that transactions in a blockchain can be viewed as elements within a database. Therefore, a mempool containing a total of $N$ transactions, of which $M$ are tampered, can be associated with an $N$-element database. The search problem of tampered transactions is then reduced to identifying the $M$ elements within the set $\mathcal{M}$ of tampered transactions in the database. Moreover, as mentioned in Section \ref{subsec:graph_representation}, database elements can be represented as nodes in the transaction graph. 

Although a search algorithm with multiple marked elements, such as the generalized Grover algorithm \cite{Grover_M}, could be used as a candidate, a ranking algorithm like PageRank offers additional advantages. As mentioned in Section \ref{subsec:Classical_PageRank}, the PageRank algorithm ranks pages by considering the number of incoming links they receive, with pages receiving numerous incoming links from sources that have relatively few outgoing links being more likely to achieve a higher rank. From a transaction graph perspective, attacks that manipulate multiple transactions to redirect their unspent transaction outputs (UTXOs) in order to accumulate large amounts of funds result in a node with multiple independent incoming connections in the transaction graph. Consequently, such nodes are likely to receive a high PageRank score, facilitating their identification. Furthermore, both the affected nodes and those receiving funds from the fraudulent node will be connected to it. Therefore, once a transaction of this nature is identified, it can be traced back to uncover other tampered transactions associated with it or to identify the transactions that have been impacted. This approach allows a classical search acting in parallel to be redirected in a way that enables faster detection of potential fraudulent or affected transactions, compared to an unranked algorithm such as Grover's. As a result, it provides an efficient search strategy for detecting fraudulent transactions and accelerates the process of assessing the extent of the impact on other transactions. For that reason, quantum search algorithms based on the PageRank algorithm, such as Quantum SearchRank or Randomized SearchRank, constitute more suitable candidates for identifying tampered transactions in the blockchain. Moreover, their graph-based structure naturally aligns with the transaction graph in UTXO-based blockchains, such as Bitcoin, as previously discussed. Representing database elements—transactions—as nodes in the transaction graph enables their validation through a quantum walk on the graph. 

Quantum search algorithms require the use of an oracle operator. In the context of blockchain, this oracle acts on transactions to indicate whether the transactions have been manipulated or not. Such oracle can be implemented in the blockchain considering the following. As commented in Section \ref{subsec:Quantum_SearchRank}, the function $f(x)$, utilized by the operator $Q_{f}$, can be designed to retrieve data from a database associated with the node $x$ in the transaction graph. In this context, the node represents a transaction. Consequently, a function can be modeled to read the transaction information, such as the hashed message, the digital signature, and the public key of the signatory, and return a boolean value indicating whether the transaction has been tampered with. Essentially, this is the role played by the verification function $V$, see Table \ref{tab:Cryptograpic_functions}, which is associated with the verification algorithm implemented in the blockchain. Therefore, it is essential to understand the details of the validation process of the blockchain on which we intend to apply the search algorithm.

After this digression, the Quantum Signature Validation Algorithm can be defined as follows:
A transaction graph is constructed from the transactions in the mempool of a validator, as described in Section \ref{sec:Blockchain}. This process is straightforward for UTXO-based models, such as Bitcoin. Then, a quantum counting algorithm \cite{Q_counting_1,Q_counting_2}, which combines various quantum techniques such as Quantum Phase Estimation (QPE) and Grover's iteration with the oracle $Q_f$ of interest, is applied to the $N$-element database of transactions that constitute the mempool in order to determine the total number of manipulated transactions in the database, $M$. Next, a PageRank-based quantum search algorithm, such as the algorithms explored in Section \ref{subsec:Quantum_SearchRank} and Section \ref{subsec:Randomized_SearchRank}, is performed on the transaction graph. Once the quantum walk is completed, the quantum algorithm will produce a probability distribution associated with each transaction in the mempool. After performing the measurement process at time $t=\lfloor\sqrt{N/M} \hspace{1.5pt}\rceil$, the highest-ranked tampered transaction is expected to be obtained. Subsequently, the resulting transaction is registered as identified by modifying the oracle with a filter that reinverts its phase. Additionally, a classical search algorithm can be applied to the transactions linked to the identified one in order to analyze the impact or the extent of the fraud. Finally, the system state is reset to its initial state for the next iteration of the algorithm. The process is executed at most $M$ times until all falsified transactions have been identified. This algorithm amplifies the probability of measuring first the fraudulent transactions with multiple unspent fund inflows coming from other transactions, which cause a significant and dispersed impact on legitimate transactions. Additionally, it facilitates an efficient search strategy for detecting fraudulent or impacted transactions, as explained above.

As a consequence of the added time complexity of the Quantum Counting algorithm, the Quantum Signature Validation Algorithm identifies corrupted transactions in an expected time of $O(\sqrt{N})$. Nevertheless, this still represents a quadratic speedup compared to its classical equivalent. Moreover, the QSVA provides a hierarchical structure of fraudulent transactions that allows tracking their impact on other transactions and offers an efficient search strategy.

A summary of the steps involved in the QSVA is presented below:

\vspace{1.75mm}

\begin{center}
\noindent {\em \bf Quantum Signature Validation Algorithm}
\end{center}

\begin{figure*}[hbtp]
    \includegraphics[scale=0.6]{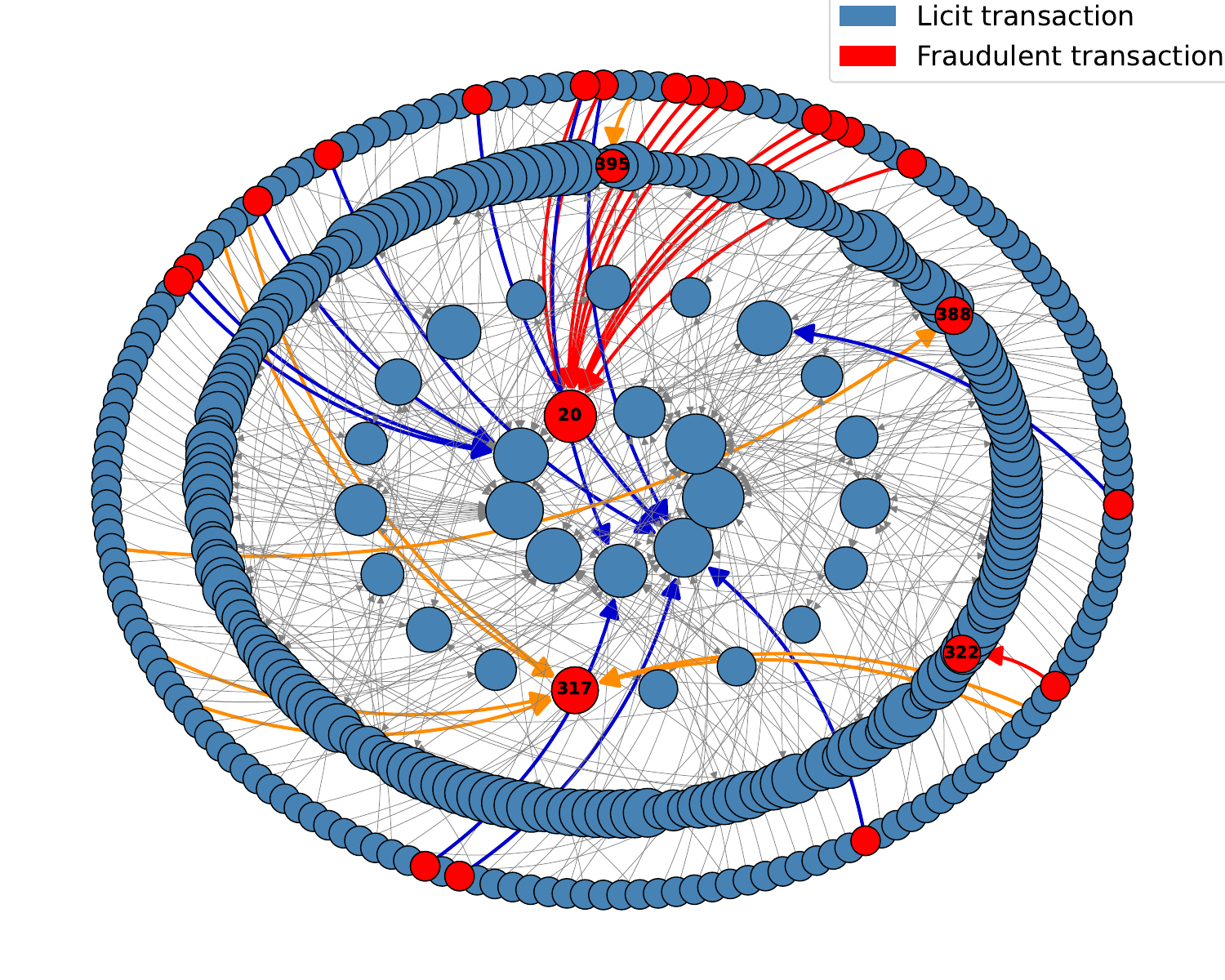}
     \caption{Transaction graph of the filtered dataset. Fraudulent transactions are represented in red, while licit transactions are shown in blue. The nodes in the graph are arranged in concentric circles based on their in-degree. Nodes with an in-degree of 0 are positioned in the outermost circle; nodes in the second and third circles have in-degrees of 1 and between 2 and 9, respectively; and the innermost circle consists of nodes with an in-degree of 10 or higher. The size of each node is proportional to its Classical PageRank value, scaled logarithmically. Edges are color-coded based on the type of transactions they connect: blue edges represent connections from a fraudulent transaction (origin) to a licit transaction (destination); orange edges represent connections from a licit transaction (origin) to a fraudulent transaction (destination); and red edges indicate connections where both the origin and destination are fraudulent transactions. Finally, the top five fraudulent transactions listed in Table \ref{tab:Fraudulent_table_t37} are labeled in the graph.}
     \label{fig:Graph_fraudulent}
\end{figure*}

\begin{description}
\item[Step 1/] Construct a transaction graph from the transactions in the mempool of a validator: nodes represent transactions and directed edges the relationships between these transactions.

\item[Step 2/] Apply a quantum counting algorithm to the $N$-element database of transactions that constitute the mempool to obtain the total count of manipulated transactions in the database, $M$.

\item[Step 3/] Perform a PageRank-based quantum search algorithm on the transaction graph from Step 1, measuring the system at time $t=\lfloor\sqrt{N/M} \hspace{1.5pt}\rceil$, to rank tampered transactions based on their connectivity. 

\item[Step 4/] Register the tampered transaction obtained from the measurement process in Step 3 as identified by adjusting the filter in the oracle of the PageRank-based quantum search algorithm.

\item[Step 5/] Evaluate the possibility of conducting a classical search on transactions linked to the identified fraudulent transaction to detect additional potentially tampered or affected transactions.

\item[Step 6/] Repeat Steps 3, 4, and 5 at most $M-1$ times to identify all tampered transactions in the mempool. 
\end{description}

\section{Simulation Results}\label{sec:Simulation_results}

Given that current quantum computers lack the capacity and development necessary to effectively perform a simulation of this nature, we will evaluate the functionality of the Quantum Signature Validation Algorithm using SQUWALS \cite{Squwals}, a quantum walk simulator implemented on a classical computer. From a dataset of transactions, serving as a database of elements, SQUWALS enables the efficient simulation of the operators involved in the Szegedy quantum walk given the Google matrix $G$. Moreover, the simulator can emulate the action of the oracle $Q_f$ on the transactions using the expression \eqref{eq:oracle_graph}. This oracle simulates the transaction validation process, analogous to the function $V$, which involves analyzing the transaction's hashed message, digital signature, and the public key of the signatory to determine whether the transaction is tampered or not. In order to evaluate the performance of the algorithm, we have selected a dataset consisting of real transactions extracted from the Bitcoin blockchain. This dataset includes a subset of transactions that have been previously classified as fraudulent \cite{Kaggle}. The primary objective of the algorithm is to accurately identify all the falsified transactions within this dataset, demonstrating its effectiveness in detecting fraudulent activity in blockchain networks. 

The transactions dataset employed is called the Elliptic Dataset, which is a publicly available Bitcoin dataset provided by Elliptic, a blockchain analytics company designed for research on detecting fraudulent activity in cryptocurrency transactions \cite{Elliptic}. The Elliptic dataset is hosted on Kaggle \cite{Kaggle}, where it is freely available for download. The dataset comprises three distinct types of data: information related to each transaction, including the transaction identifier, the timestamp when the transaction was recorded, and the amount of BTC transferred; the classification of each transaction as either licit, fraudulent, or unclassified; and the origin and destination transactions associated with the flow of funds. We filtered the dataset to include only licit and fraudulent transactions recorded within a specific time frame, simulating the mempool of transactions a validator would encounter at a given moment, with fraudulent transactions treated as tampered. 

From the filtered dataset of classified transactions, which includes the origin and destination transactions associated with BTC movements, we constructed the corresponding transaction graph, similar to the one shown in Figure \ref{fig:UTXO_graph}. This graph serves as the basis for applying the graph-based algorithms described in Section \ref{sec:Algorithms}, including Classical PageRank, Quantum SearchRank, and Randomized SearchRank. Figure \ref{fig:Graph_fraudulent} illustrates the transaction graph derived from the filtered dataset, highlighting fraudulent transactions and their associated money flows in colors. The size of the nodes representing transactions is determined based on their PageRank score, scaled logarithmically. 

\begin{figure}[hbtp]
    \centering
    \includegraphics[width=0.95\linewidth]{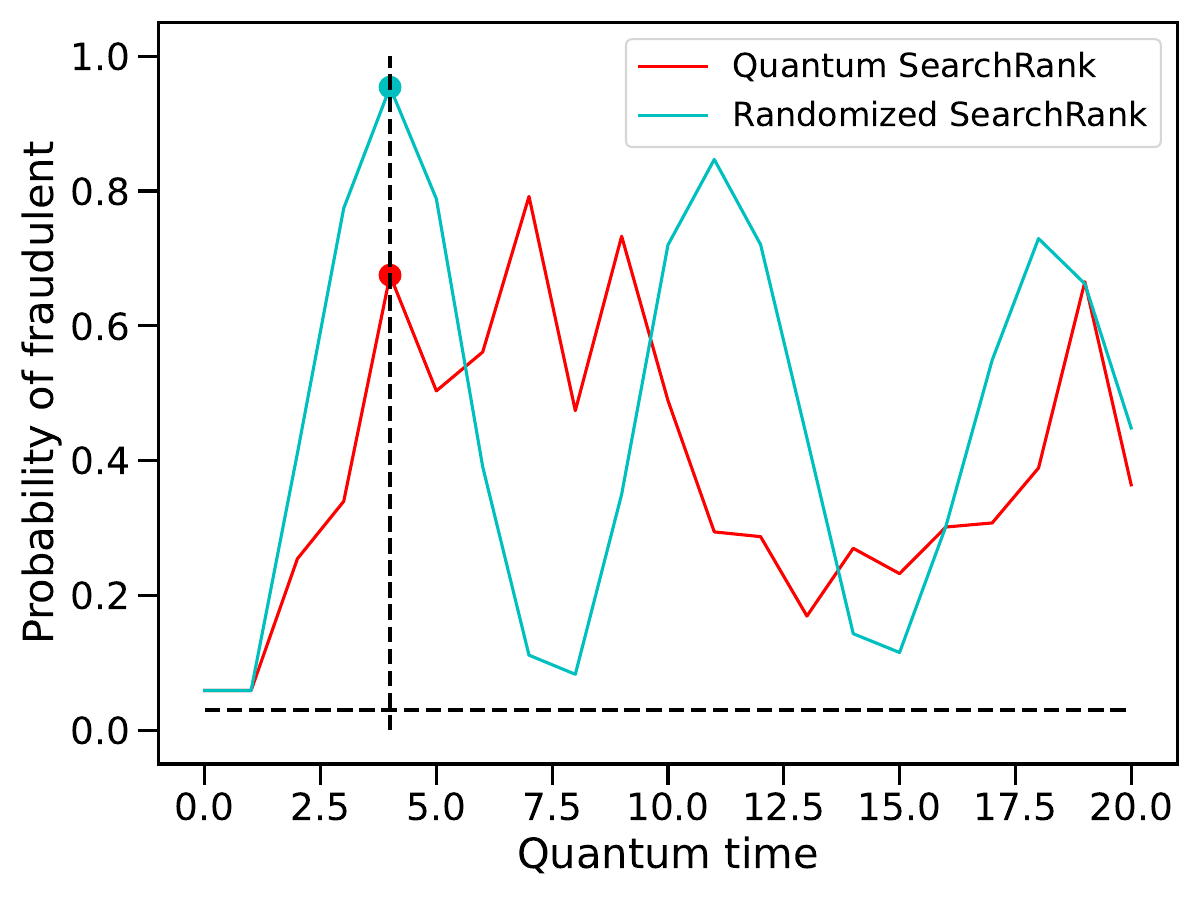}
    \caption{Probability of measuring a fraudulent transaction during the first iteration of the QSVA as a function of the number of steps in which the unitary evolution operator is applied. The first maximum of each curve is marked with a dot. The vertical dashed line represents the reference time $t=\lfloor\sqrt{N/M} \hspace{1.5pt}\rceil$. The horizontal dashed line represents the probability of the fraudulent transactions in the Classical PageRank distribution.}
    \label{fig:Time_i_1}
\end{figure}

We have tested both quantum search algorithms, Quantum SearchRank and Randomized SearchRank, on the transaction graph to determine whether either of these graph-based algorithms could serve as a viable candidate for the Quantum Signature Validation Algorithm, as both are built upon the PageRank algorithm at their core. 

\begin{figure}[hbtp]
	\centering
	\subfigure[]{\includegraphics[width=\linewidth]{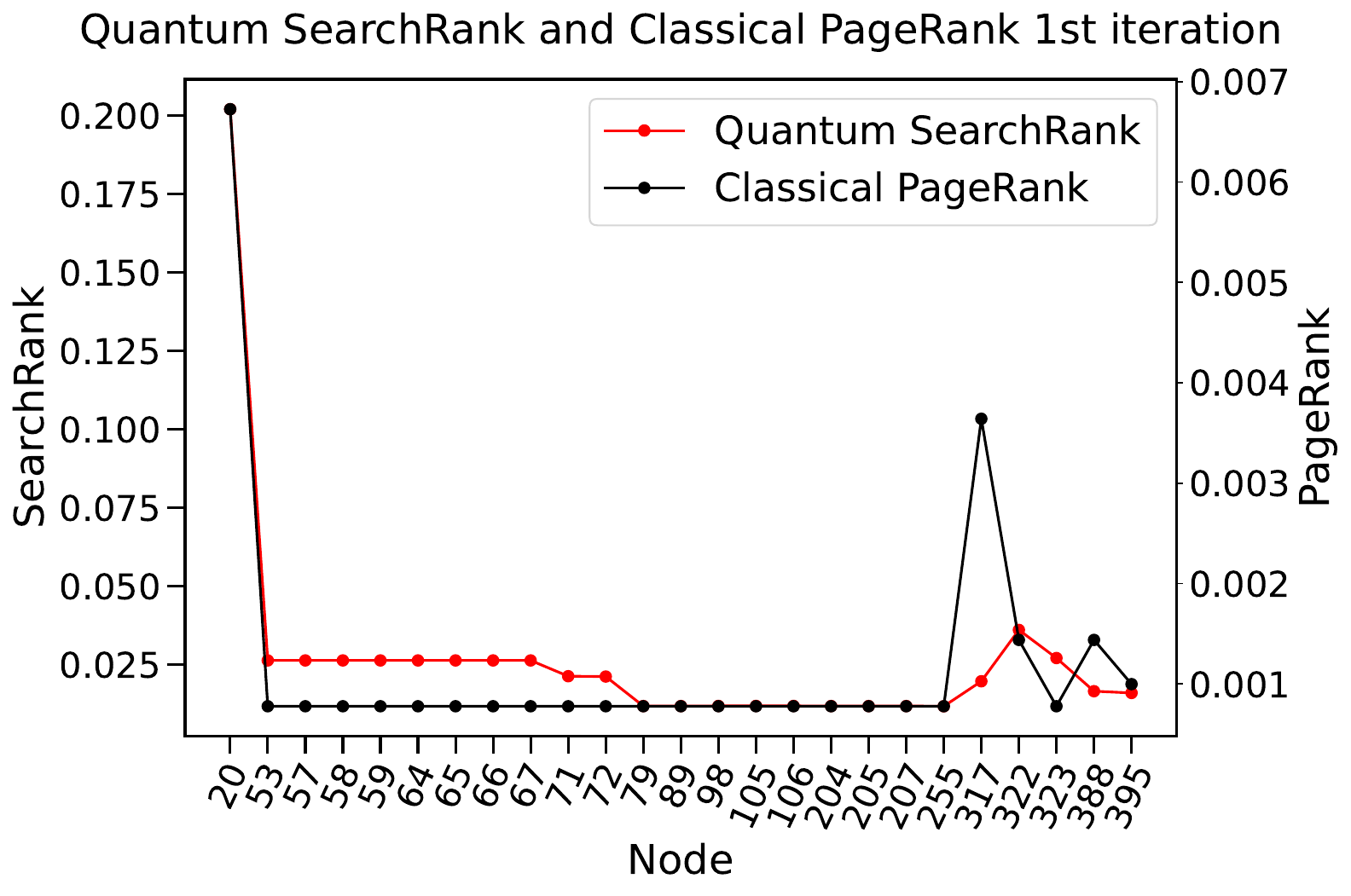}}\label{fig:QS_i_1}
	\subfigure[]{\includegraphics[width=\linewidth]{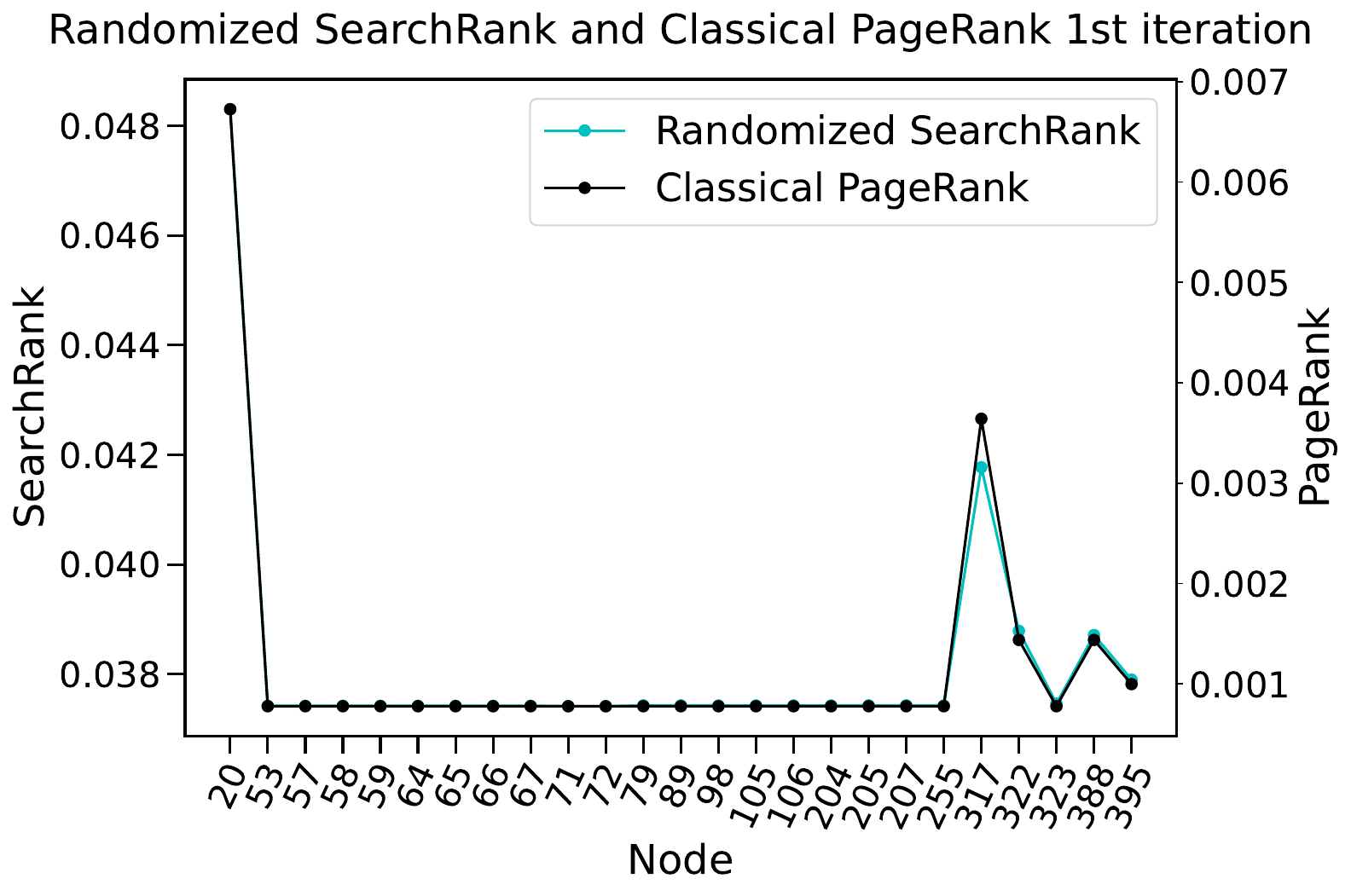}}\label{fig:QS_i_1}
        \caption {a) Probability distribution of tampered transactions for the Quantum SearchRank and Classical PageRank during the first iteration of the QSVA.
                b) Probability distribution of tampered transactions for the Randomized SearchRank and Classical PageRank during the first iteration of the QSVA. The PageRank (right axis) is represented on a different scale from the SearchRank (left axis).}
	\label{fig:First_iteration}
\end{figure}

Following the procedure outlined in Section \ref{subsec:QSVA}, we first computed the probability distributions associated with measuring each tampered transaction for all time steps, and then selected the one corresponding to $t=\lfloor\sqrt{N/M} \hspace{1.5pt}\rceil$ for each iteration of the algorithm. Figure \ref{fig:Time_i_1} illustrates the probability of measuring a fraudulent transaction as a function of the number of applications of the unitary evolution operator, as defined in Equation \eqref{eq:unitary_evolution}, during the first iteration of the QSVA. It is noteworthy that the probability reaches its first maximum at $t=\lfloor\sqrt{N/M} \hspace{1.5pt}\rceil$ for both SearchRank algorithms. However, while the probability is maximized for the Randomized SearchRank, this is not the case for the Quantum SearchRank. The probability distribution of fraudulent transactions resulting for both quantum search algorithms after the first iteration of the QSVA are shown in Figure \ref{fig:First_iteration}. Since the probability of measuring a fraudulent transaction has been amplified in the SearchRank algorithms, we represent the SearchRank distributions on a different scale than the PageRank distributions. Notably, the Randomized SearchRank distribution closely aligns with the Classical PageRank distribution, in contrast to the Quantum SearchRank. It is particularly striking that several transactions exhibit identical scores in the Classical PageRank, highlighting the prevailing patterns of connections among transactions within the network. Once again, the Randomized SearchRank proves to be a better fit for these degenerate transactions. However, the differences between their probabilities in the Randomized SearchRank are so minimal that they may also be regarded as degenerate.

\setcounter{figure}{8}

\begin{figure}[hbtp]
    \centering
    \includegraphics[width=0.95\linewidth]{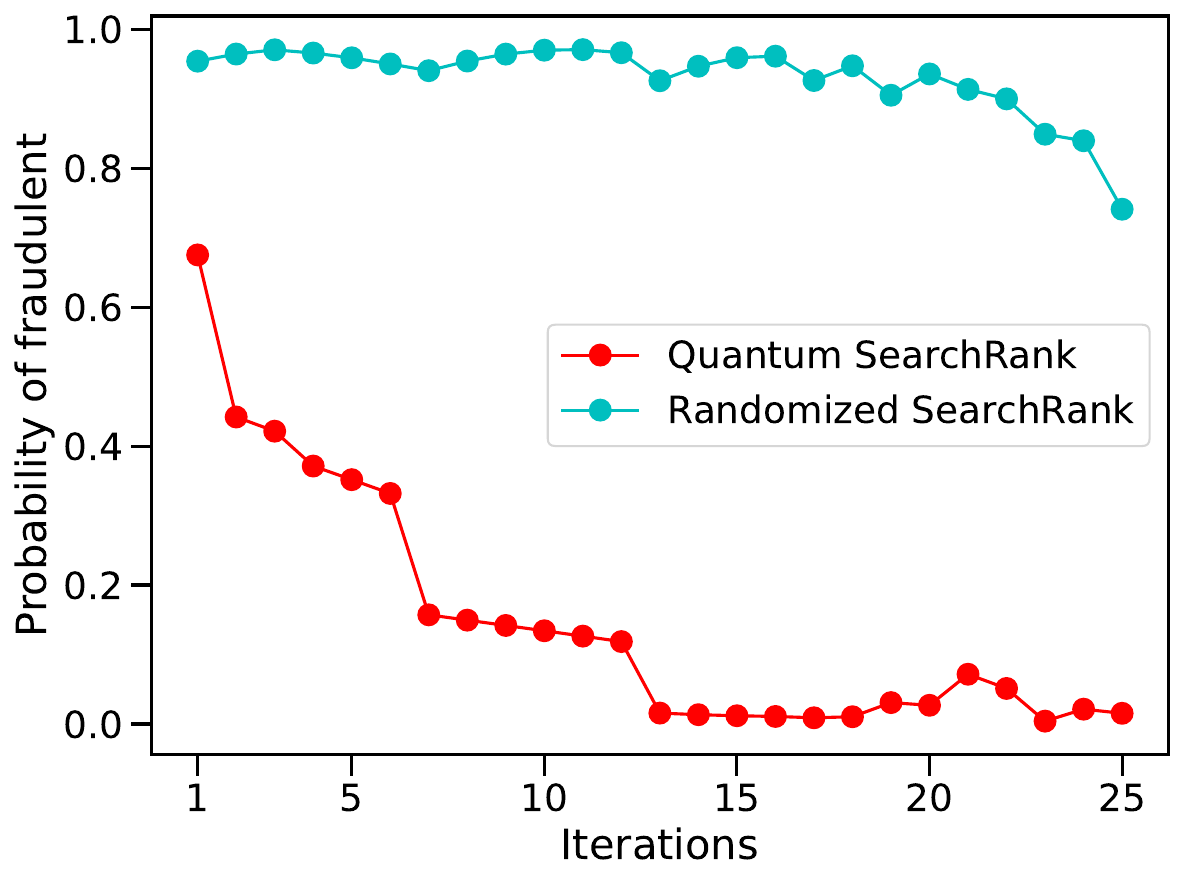}
    \caption{Probability of measuring a fraudulent transaction at the reference time $t=\lfloor\sqrt{N/M} \hspace{1.5pt}\rceil$ as a function of the QSVA iterations.}
    \label{fig:probability_vs_iteration}
\end{figure}

Adhering to the QSVA procedure, the transaction most likely to be obtained after measurement is the one associated with node 20, see Figure \ref{fig:First_iteration}. Based on this result, one can trace the transactions connected to it to uncover ten additional fraudulent transactions with a straightforward classical verification, see Figure \ref{fig:Graph_fraudulent}. Therefore, less iterations of the QSVA procedure would be needed in the end. In contrast, an unranked quantum search algorithm such as Grover's, without any structured prioritization, would detect whatever fraudulent node first, so it would be helpless for a parallel classical search. Thus, with Grover's algorithm, one would need to perform $M$ iterations of the algorithm to detect all the marked nodes, making the whole process less efficient.

For simulations purposes we consider that QSVA is performing alone, so we keep iterating the algorithm until finding the $M$ marked nodes. The probability distributions of the consecutive four iterations are illustrated in Figure \ref{fig:superfig_t37}. Given the closer alignment of the Randomized SearchRank with the Classical PageRank distribution compared to the Quantum SearchRank, the most likely transactions expected to be obtained after measurement in each iteration of the QSVA coincide with the high-ranked transactions of the Classical PageRank when considering the Randomized SearchRank.
In addition, Figure \ref{fig:probability_vs_iteration} shows the probability of measuring a tampered transaction as a function of the QSVA iterations for each quantum algorithm. Note that the probability of measuring a fraudulent transaction decreases with each iteration of the QSVA when using the Quantum SearchRank, whereas it remains close to one when using the Randomized SearchRank. The figure illustrates that the Quantum SearchRank fails to effectively amplify the probability of measuring a fraudulent transaction as QSVA iterations progress. For the second iteration, the probability of measuring a fraudulent transaction drops below $50\%$ for the Quantum SearchRank, while it remains above $95\%$ for the Randomized SearchRank. By the final iteration, the probability decreases to slightly more than $1.5\%$ for the Quantum SearchRank, while it remains slightly less than $75\%$ for the Randomized SearchRank. As discussed in Section \ref{subsec:Randomized_SearchRank}, this sharp decline in the Quantum SearchRank is caused by the ratio $N/M$, which increases as the number of marked elements ($M$) is reduced when fraudulent transactions are identified. In contrast, the Randomized SearchRank is largely independent of this ratio, allowing the probability to remain consistently high. Therefore, the QSVA based on the Randomized SearchRank is able to accurately identify all the fraudulent transactions in the network. 

The first five tampered transactions expected to be observed after measurement using the QSVA are summarized in Table \ref{tab:Fraudulent_table_t37}. These transactions are listed in the order of their PageRank scores. As previously discussed, this order aligns with the most probable transactions identified by the Randomized SearchRank at each iteration. The remaining twenty fraudulent transactions are considered degenerate for the Classical PageRank and the Randomized SearchRank. Therefore, their order of measurement does not provide any additional information compared to an unranked quantum search algorithm.

\setcounter{figure}{7}

\begin{figure*}[hbtp]
        \vspace{-20pt}
	\makebox[0pt][c]{
		\subfigure[]{\includegraphics[scale=0.33]{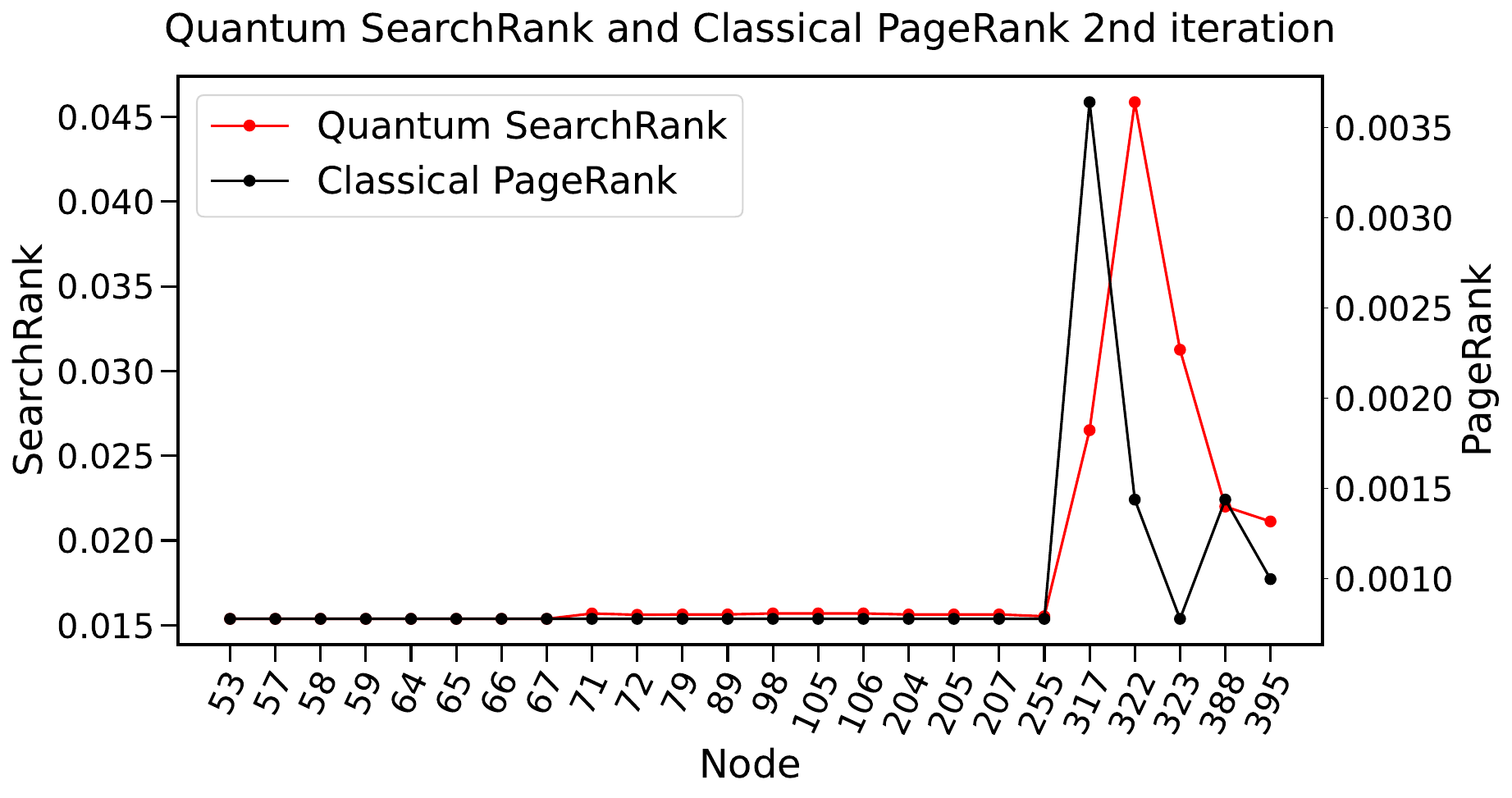}}
		\hspace{-6pt}
		\subfigure[]{\includegraphics[scale=0.33]{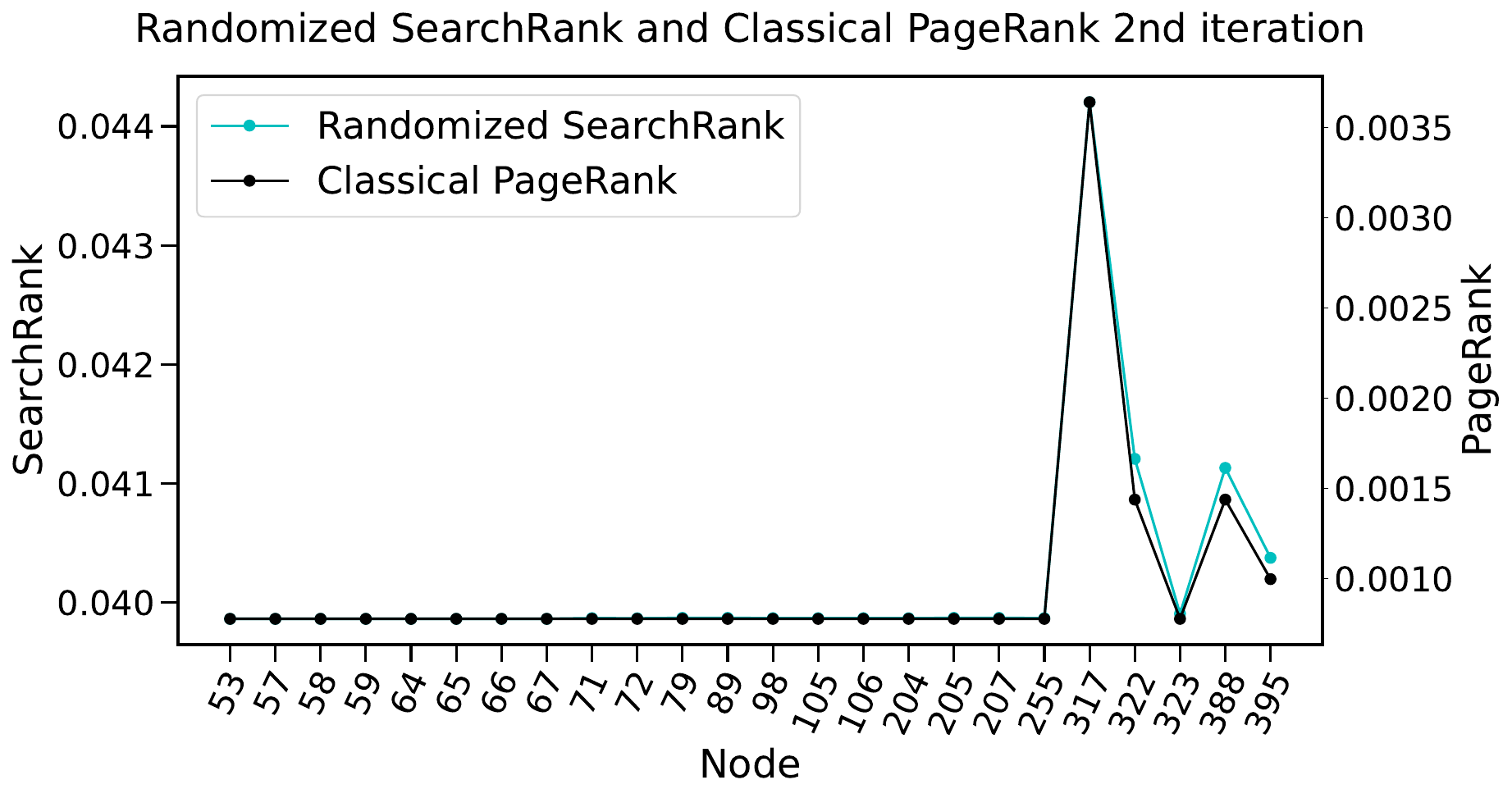}}
	}
	\\
	\vspace{-7pt}
	\makebox[0pt][c]{
		\subfigure[]{\includegraphics[scale=0.33]{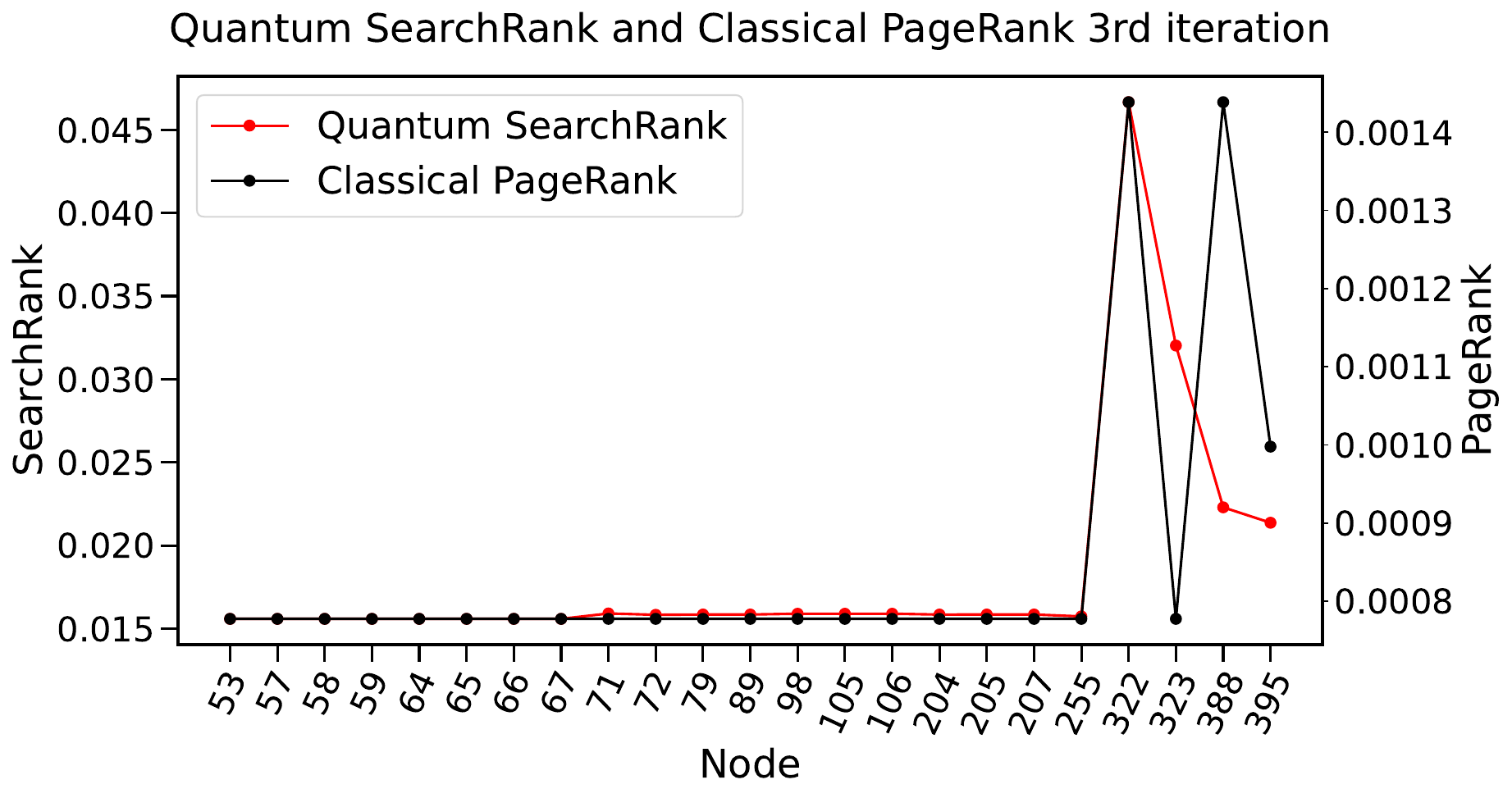}}
		\hspace{-6pt}
		\subfigure[]{\includegraphics[scale=0.33]{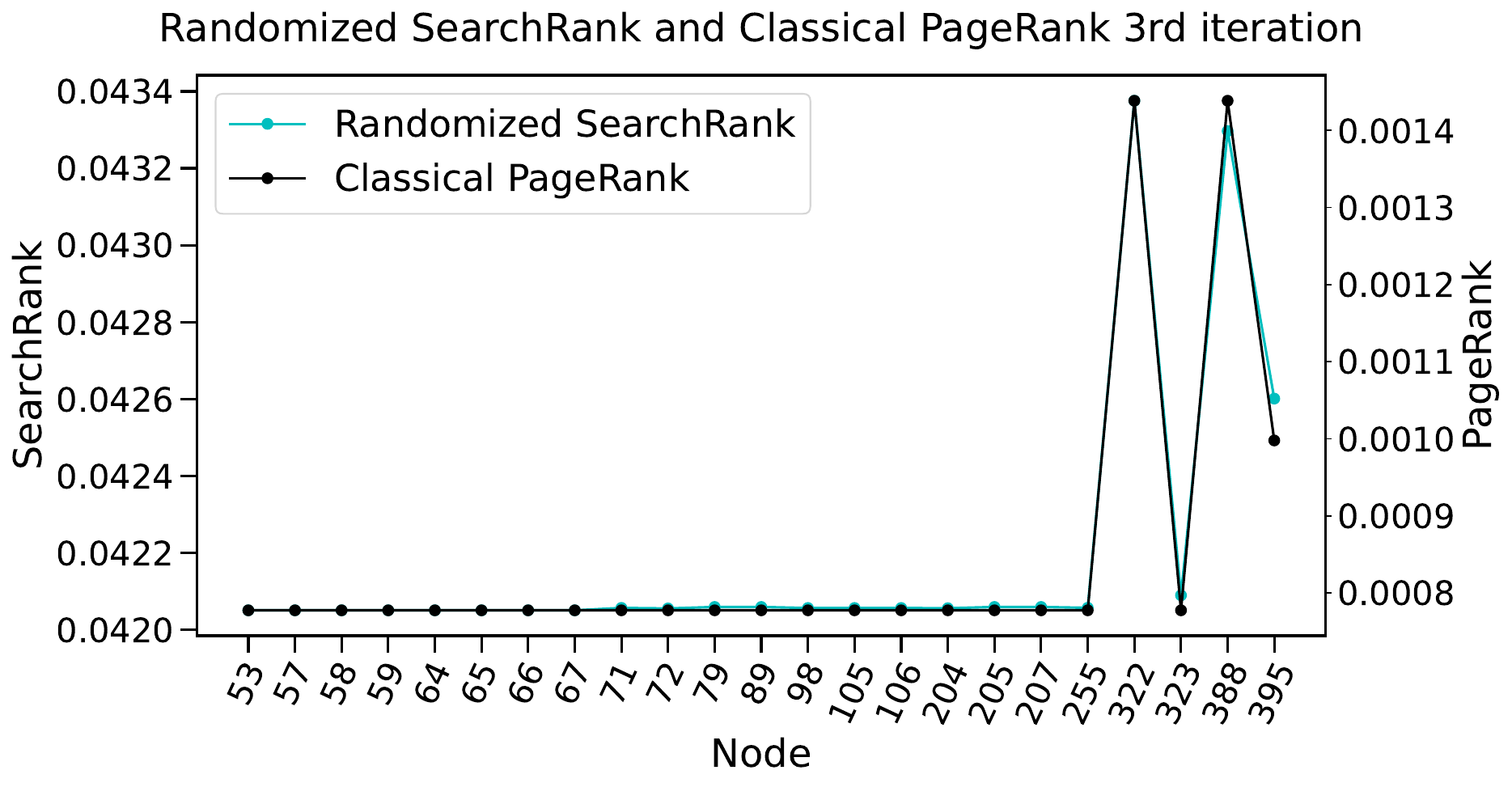}}
	}
	\\
	\vspace{-7pt}
	\makebox[0pt][c]{
		\subfigure[]{\includegraphics[scale=0.33]{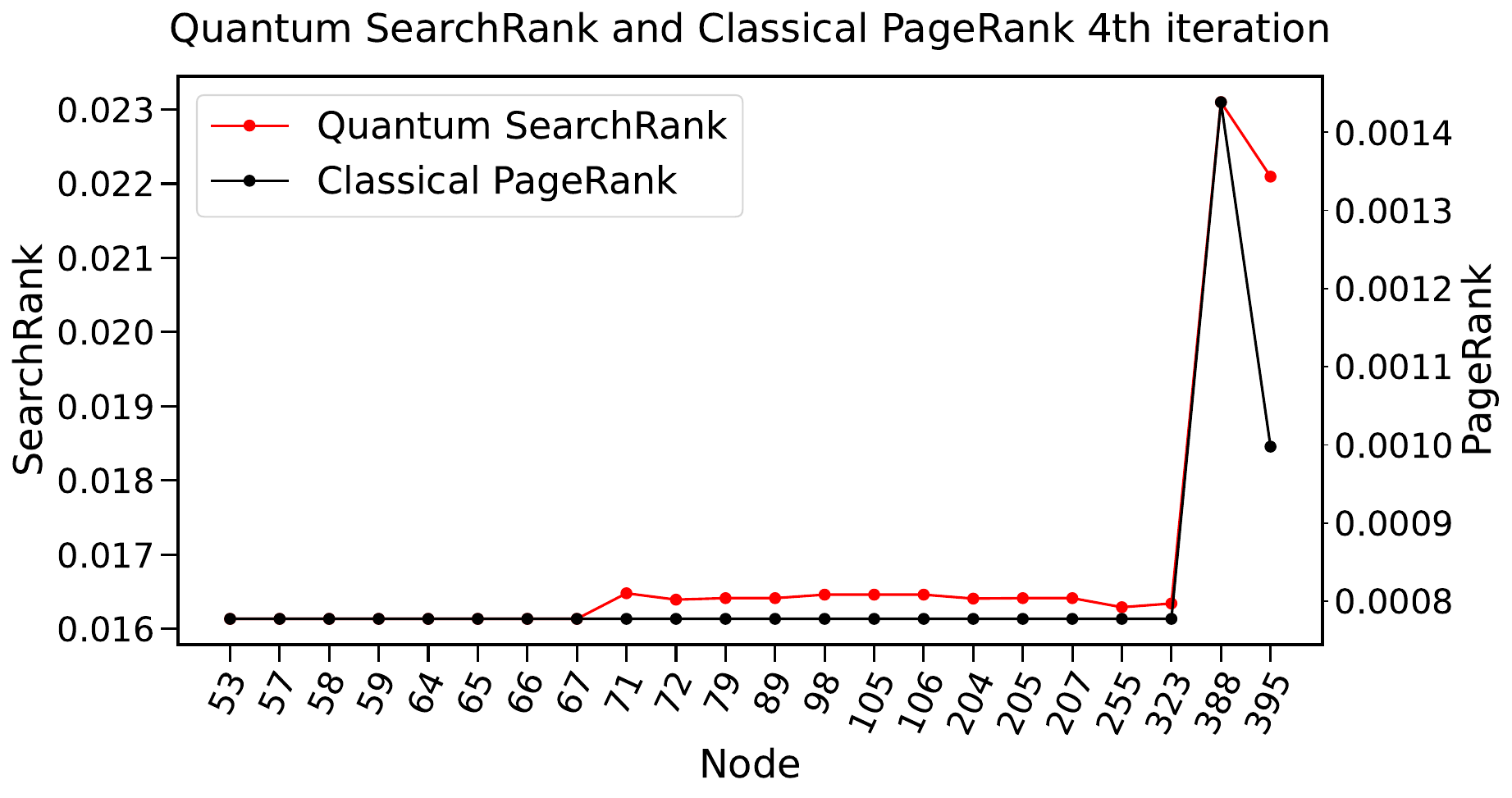}}
		\hspace{-6pt}
		\subfigure[]{\includegraphics[scale=0.33]{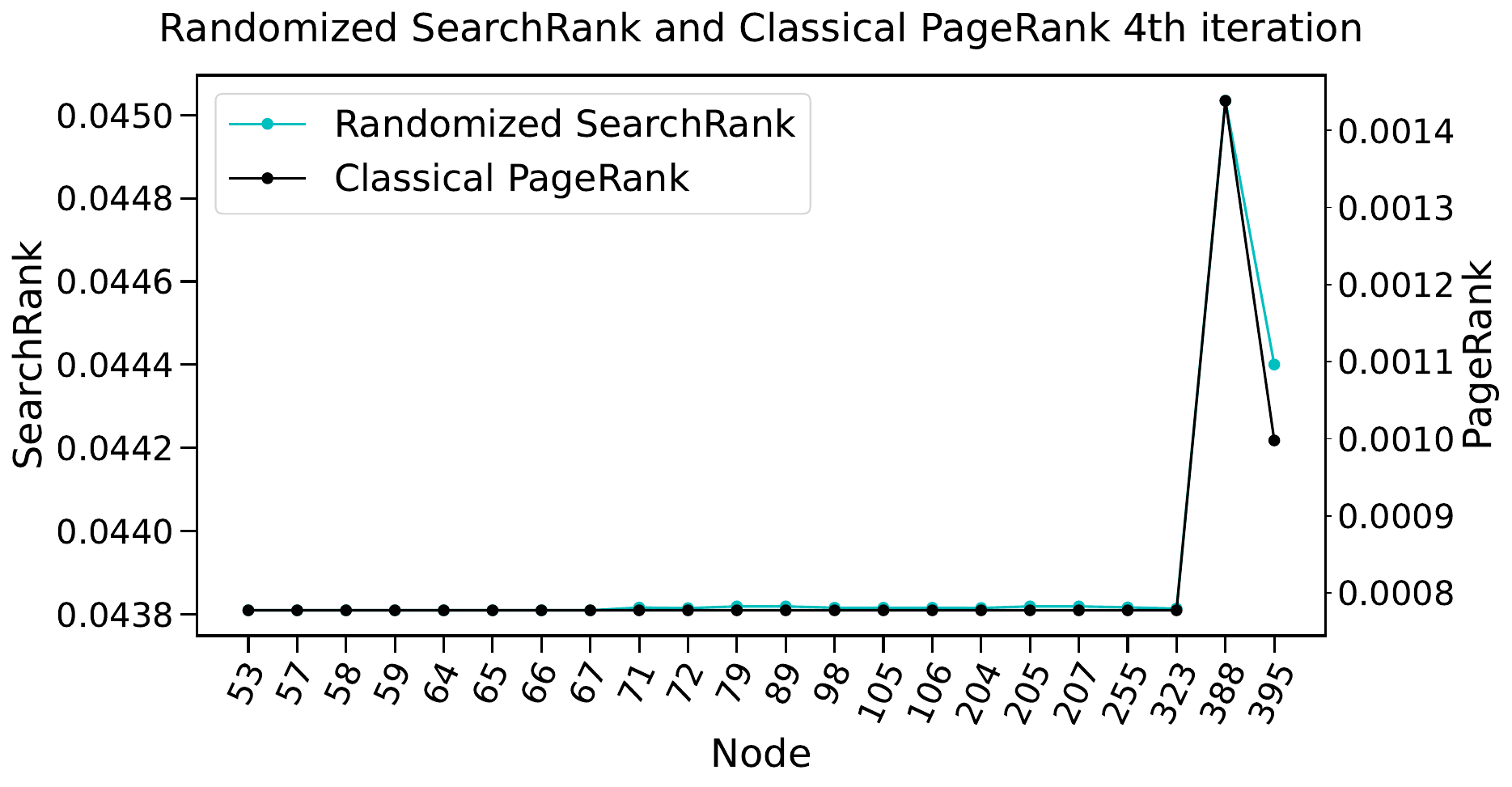}}
        }
	\\
	\vspace{-7pt}
	\makebox[0pt][c]{
		\subfigure[]{\includegraphics[scale=0.33]{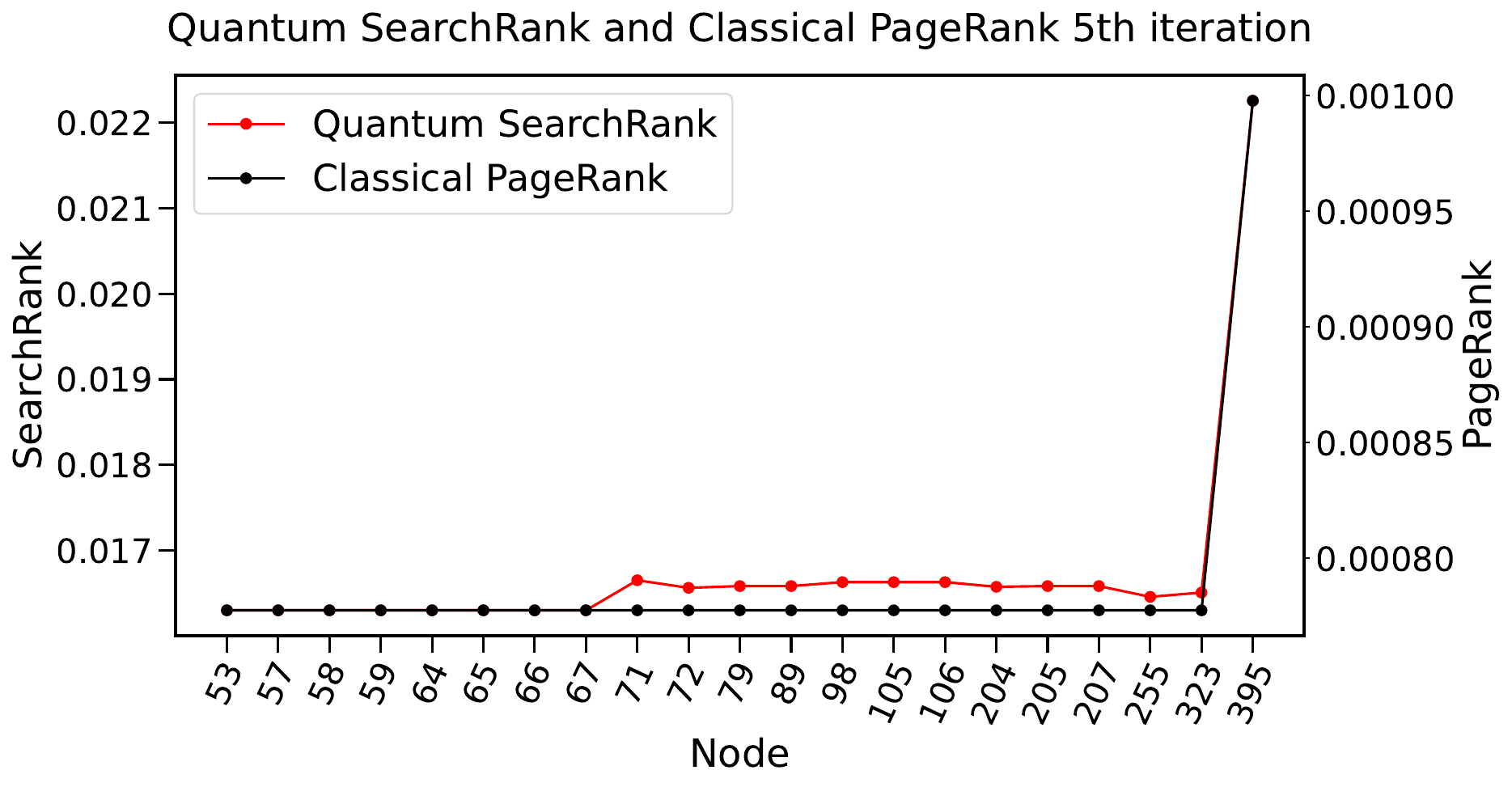}}
		\hspace{-6pt}
		\subfigure[]{\includegraphics[scale=0.33]{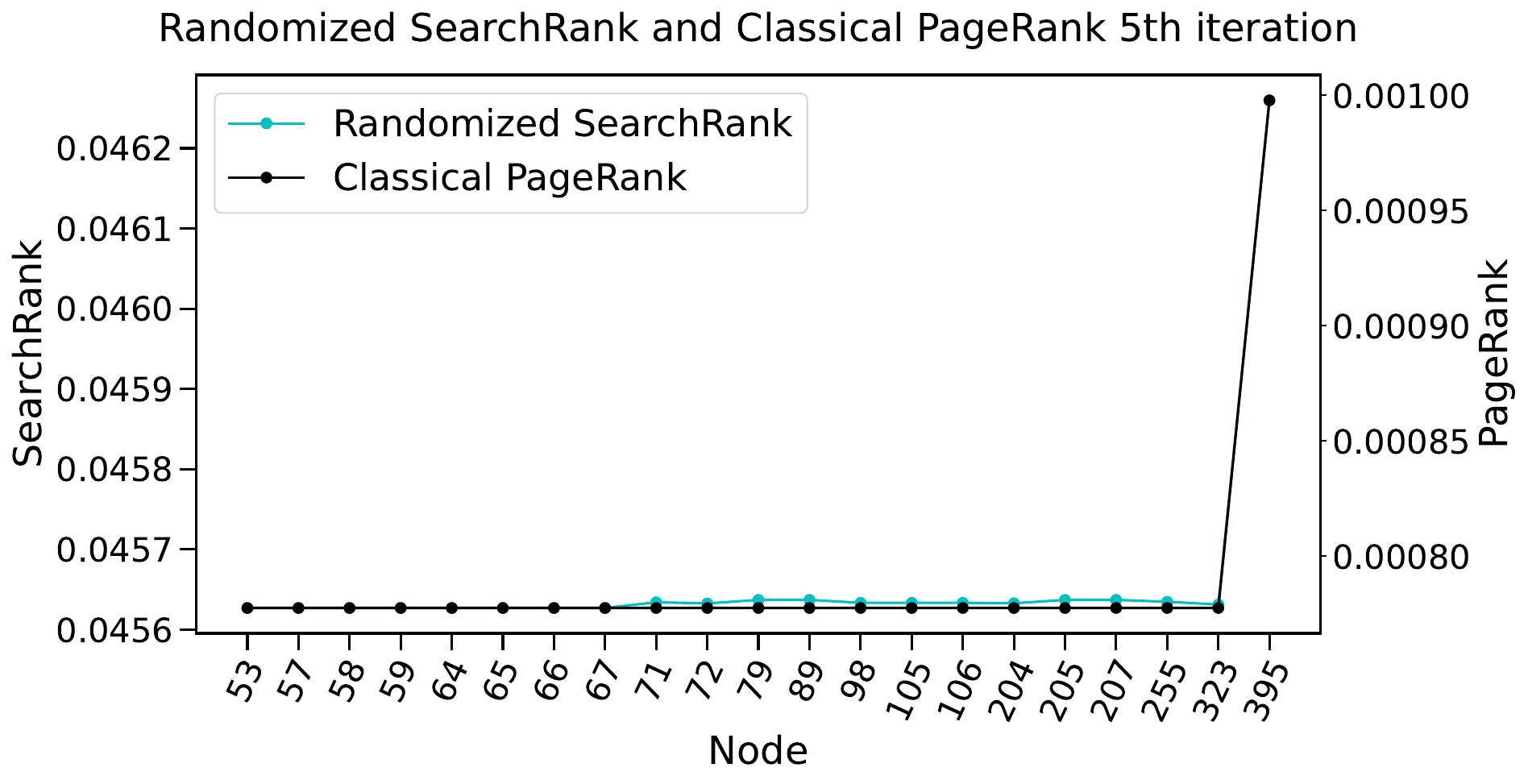}}
	}
	\caption{Representation of iterations 2, 3, 4 and 5 of the Quantum Signature Validation Algorithm. For each iteration, the left panel illustrates the probability distribution generated by the Quantum SearchRank and the Classical PageRank for each node. Similarly, the right panel displays the probability distribution produced by the Randomized SearchRank and the Classical PageRank for each node. The PageRank (right axis) is represented on a different scale from the SearchRank (left axis).}
	\label{fig:superfig_t37}
\end{figure*}

\setcounter{figure}{9}

\begin{table}[hbtp]
\caption{The top five fraudulent transaction identifiers and their corresponding graph node numbers, listed in descending order of their PageRank scores. This order reflects the sequence in which they are expected to be identified after measurement.}
\centering
\begin{tabular}{lc}
\textbf{Node} & \textbf{TxId} \\ \hline
\addlinespace[8pt]
20 & 13735016 \\
317 & 30179316 \\
322 & 30204549 \\
388 & 39684200 \\
395 & 39747107
\end{tabular}
\label{tab:Fraudulent_table_t37}
\end{table}

\section{Conclusions}\label{sec:Conclusions}

We have reviewed the fundamentals of blockchain technology, focusing on the mechanisms of transaction validation within transaction-based blockchains. Additionally, we introduced the transaction graph representation of blockchain systems, where nodes represent individual transactions, and directed edges connect transactions referencing unspent outputs. This representation is particularly intuitive for UTXO-based models, such as Bitcoin, and facilitates efficient transaction verification. Unlike account-based models, it avoids the need to track the complete transaction history of an address to confirm sufficient funds. Instead, verification relies on maintaining an up-to-date set of unspent transaction outputs (UTXOs), which ensures accurate and efficient validation. 

We introduced the Classical PageRank algorithm as a powerful tool for extracting meaningful insights from transaction graphs. This algorithm ranks nodes based on the principle that nodes receiving numerous incoming links from sources with relatively few outgoing links are more likely to achieve a higher rank. This ranking is particularly relevant for detecting attacks that involve numerous minor, independent transactions, which create a dispersed impact on the network. The PageRank algorithm can thus guide parallel search algorithms toward identifying fraudulent transactions linked to a tampered transaction or affected by it. Building on this foundation, we reviewed the principles and formulation of two PageRank-based quantum search algorithms: Quantum SearchRank and Randomized SearchRank. These quantum algorithms demonstrate a quadratic speedup over their classical counterparts in element searching, offering a promising approach for the rapid identification of fraudulent or impacted transactions. Moreover, we have compared both algorithms, highlighting their respective scopes and limitations.

Since identifying fraudulent transactions promptly allows their removal from the mempool and replacement with valid transactions to be incorporated into a block for execution, quantum algorithms could significantly enhance the detection speed of tampered transactions, thereby improving overall blockchain efficiency. To leverage the advantages of the quantum algorithms presented while preserving the search strategy provided by the PageRank algorithm, we propose the Quantum Signature Validation Algorithm (QSVA). The QSVA utilizes the transaction-graph representation to validate transactions through a quantum walk performed using a PageRank-based search algorithm. Its PageRank core, in conjunction with a classical search, redirects the search process to enable faster detection of potentially fraudulent or impacted transactions compared to unranked algorithms, such as Grover’s. This approach not only provides an efficient search strategy but also benefits from the quantum algorithm's quadratic speedup, significantly reducing the time required to identify tampered transactions within the blockchain to $O(\sqrt{N})$, being N the number of transactions of the dataset, when compared to classical search algorithms. 

Due to the limitations of current quantum computers, we evaluated the functionality of the QSVA using a quantum walk simulator implemented on a classical computer. For this purpose, we employed the SQUWALS simulator, which emulates the action of the oracle in a quantum search algorithm, assuming a known set of manipulated transactions. To examine its performance in a realistic scenario, we utilized a publicly available Bitcoin transaction dataset containing real transactions previously classified as fraudulent or legitimate. This dataset simulates the mempool that an average validator would encounter while compiling a block of transactions in a blockchain. 

Additionally, we tested whether the Quantum SearchRank or the Randomized SearchRank could serve as viable candidates for the QSVA. The results show that the QSVA, when executed with either the Quantum or Randomized SearchRank, amplifies the probability of detecting a fraudulent transaction from the very first iteration of the algorithm. Moreover, the primary objective of the QSVA—successfully identifying all fraudulent transactions within the dataset in less time than a classical algorithm—was achieved in simulations employing the Randomized SearchRank. Notably, the Randomized SearchRank more accurately amplifies highly ranked transactions as determined by the Classical PageRank algorithm. In addition, we observed that the Randomized SearchRank aligns more closely with the PageRank distribution of fraudulent transactions, outperforming the Quantum SearchRank in this regard. We also pointed out that the probability of detecting a fraudulent transaction reaches its maximum at $t=\lfloor\sqrt{N/M} \hspace{1.5pt}\rceil$ for the Randomized SearchRank, aligning with the optimal runtime of a quantum search algorithm. Furthermore, we highlighted a critical limitation of the Quantum SearchRank: its scalability. As iterations of the QSVA progress, the probability of detecting a fraudulent transaction decreases significantly, reducing its effectiveness. Conversely, the Randomized SearchRank maintains a consistently high probability of detecting falsified transactions. Based on these findings, we conclude that the Randomized SearchRank is a more suitable candidate for the QSVA than the Quantum SearchRank. 

Although the results obtained for the Quantum Signature Validation Algorithm (QSVA) are promising, as is the case with most quantum algorithms proposed for blockchain, further research and development on current quantum hardware are essential to achieve high-fidelity simulations. Specifically, future work focused on the circuit design of the oracle used for transaction validation will be required for the QSVA. Additionally, advancements in the design of the operators involved in the quantum walk are necessary to enable the efficient implementation of the SearchRank algorithms, tailored to the specific structure of the transaction graph \cite{Q_circuits}. Fortunately, quantum hardware is advancing at an impressive pace, consistently overcoming previous limitations. Therefore, we remain increasingly optimistic about the transformative potential of quantum applications in this field. 

Recent advancements in Distributed Ledger Technologies (DLTs) present a promising pathway for integrating quantum algorithms beyond traditional blockchain systems. While blockchain is a prominent subset of DLTs, emerging systems like Hashgraph \cite{Hashgraph} and Holochain \cite{Holochain} offer unique architectures that could greatly benefit from quantum advancements. Hashgraph, for instance, uses a directed acyclic graph (DAG) model, which provides consensus more efficiently and with higher transaction throughput than traditional blockchain mechanisms. Quantum algorithms could further enhance these capabilities by optimizing consensus processes and improving the speed and reliability of transaction validations. Similarly, Holochain diverges from conventional blockchains by forgoing a global ledger, instead relying on a distributed peer-to-peer network where each node maintains its own chain of data. Quantum approaches could facilitate more robust validation strategies, ensuring data integrity and enhancing security in these decentralized environments. By addressing the specific needs of varying DLT architectures, quantum computing holds the potential to unlock unprecedented levels of scalability, efficiency, and security across a diverse array of distributed systems. Continued research in this area could significantly advance the integration of quantum solutions into these platforms, opening new frontiers for innovation and application in decentralized technologies.

\section*{Acknowledgments}

We acknowledge the support from the Spanish MINECO grants MINECO/FEDER Projects,  PID2021-122547NB-I00 FIS2021, the “MADQuantum-CM" project funded by Comunidad de Madrid and by the Recovery, Transformation, and Resilience Plan – Funded by the European Union - NextGenerationEU and Ministry of Economic Affairs Quantum ENIA project. This work has also been financially supported by the Ministry for Digital Transformation and of Civil Service of the Spanish Government through the QUANTUM ENIA project call – Quantum Spain project, and by the European Union through the Recovery, Transformation and Resilience Plan – NextGenerationEU within the framework of the Digital Spain 2026 Agenda. M. A. M.-D. has been partially supported by the U.S.Army Research Office through Grant No. W911NF-14-1-0103. S.A.O. acknowledges support from Universidad Complutense de Madrid - Banco Santander through Grant No. CT58/21-CT59/21.

\newpage

\bibliography{MiBiblio}
\bibliographystyle{unsrt}

\end{document}